\documentclass[12pt,a4paper]{article}
	
\usepackage{a4wide}
\usepackage{amsmath}
\usepackage{amssymb}
\usepackage{amsfonts}
\usepackage{epsfig}
\usepackage{exscale}
\usepackage{float}
\usepackage{bbm}
\usepackage[numbers,sort&compress]{natbib}

\newcommand{\Z}{{\mathbb{Z}}}
\newcommand{\R}{{\mathbb{R}}}
\newcommand{\C}{{\mathbb{C}}}

\newcommand{\1}{{\mathbbm{1}}}
\newcommand{\0}{{0}}

\newcommand{\p}{\partial}

\makeatletter
\@addtoreset{equation}{section}
\makeatother

\title{\vskip-2cm Fate of Accidental Symmetries of the Relativistic Hydrogen
Atom in a Spherical Cavity}

\author{M.\ H.\ Al-Hashimi$^{a,b}$, A.\ M.\ Shalaby$^{a,c}$, and
U.-J.\ Wiese$^{b}\footnote{UJW acknowledges the hospitality of the INT at
Washington University in Seattle, where this work was completed.}$
\footnote{Contact information: M.\ H.\ Al-Hashimi: hashimi@itp.unibe.ch,
+41 31 631 8878; A.\ Shalaby, amshalab@qu.edu.qa, +974 4403 4630;
U.-J.\ Wiese, wiese@itp.unibe.ch, +41 31 613 8504.}
\\ \\
{\small $^a$ Department of Mathematics, Statistics, and Physics} \\
{\small Qatar University, Al Tarfa, Doha 2713, Qatar} \\
{\small $^b$ Albert Einstein Center for Fundamental Physics,
Institute for Theoretical Physics} \\
{\small Bern University, Sidlerstrasse 5, CH-3012 Bern, Switzerland} \\
{\small $^c$ Physics Department, Faculty of Science, Mansoura University,
Egypt} \\ \\
{\small INT, Washington University, Seattle, U.S.A., Preprint: INT-PUB-15-009}
\\
{\small PACS numbers:}}

\begin{document}

\maketitle

\vspace{-1cm}

\begin{abstract} \normalsize

The non-relativistic hydrogen atom enjoys an accidental $SO(4)$ symmetry, that
enlarges the rotational $SO(3)$ symmetry, by extending the angular momentum
algebra with the Runge-Lenz vector. In the relativistic hydrogen atom the
accidental symmetry is partially lifted. Due to the Johnson-Lippmann operator,
which commutes with the Dirac Hamiltonian, some degeneracy remains. When the
non-relativistic hydrogen atom is put in a spherical cavity of radius $R$ with
perfectly reflecting Robin boundary conditions, characterized by a self-adjoint
extension parameter $\gamma$, in general the accidental $SO(4)$ symmetry is
lifted. However, for $R = (l+1)(l+2) a$ (where $a$ is the Bohr radius and $l$ is
the orbital angular momentum) some
degeneracy remains when $\gamma = \infty$ or $\gamma = \frac{2}{R}$. In the
relativistic case, we consider the most general spherically and parity invariant
boundary condition, which is characterized by a self-adjoint extension
parameter. In this case, the remnant accidental symmetry is always lifted in a
finite volume. We also investigate the accidental symmetry in the context of the
Pauli equation, which sheds light on the proper non-relativistic treatment
including spin. In that case, again some degeneracy remains for specific
values of $R$ and $\gamma$.

\end{abstract}

\newpage

\section{Introduction}

Already in 1873, Bertrand has shown that the $1/r$ and $r^2$ potentials are
the only spherically symmetric potentials for which all bound classical orbits
are closed \cite{Ber73}. Accidental dynamical symmetries arise in many quantum
mechanics problems \cite{McI71}, ranging from the harmonic oscillator to the
hydrogen atom and to a charged particle in a constant magnetic field. In 1935
Fock noted that the hydrogen atom possesses a ``hyper-spherical'' $SO(4)$
symmetry \cite{Foc35}, and Bargmann \cite{Bar36} showed that the generators of
the accidental symmetry are the components of the Runge-Lenz vector
\cite{Len24}. For an isotropic harmonic oscillator in $d$ dimensions, the
spatial rotational symmetry $SO(d)$ is dynamically enhanced to $SU(d)$ and for
the non-relativistic hydrogen atom it is enhanced to $SO(d+1)$. For a charged
particle in a constant magnetic field, the center of the circular cyclotron
motion plays the role of the Runge-Lenz vector \cite{Lan30,Joh49,AlH09}. In
this case, translation invariance (up to gauge transformations) disguises itself
as an accidental symmetry. In all these cases, at the classical level the
accidental symmetry implies that all bound classical orbits are closed curves,
while at the quantum level it leads to enlarged degeneracies of the energy
levels of bound states. For a charged particle in a constant magnetic field,
the corresponding Landau levels are even infinitely degenerate.

In the past, we have investigated how accidental symmetries are affected when
the geometry of the problem is modified. For example, in \cite{AlH08} we have
studied the harmonic oscillator and the hydrogen atom on a 2-d cone with deficit
angle $\delta$. For general values of $\delta$, the accidental $SU(2) = SO(3)$
symmetry is then completely lifted. However, if $\frac{\delta}{2 \pi}$ is a
rational number, some accidental degeneracies remain. Remarkably, the
degeneracies correspond to fractional (neither integer nor half-integer)
``spin'' of the accidental $SU(2)$ symmetries. This subtle effect arises
because on the cone the Runge-Lenz vector is no longer self-adjoint in the
domain of the Hamiltonian. Hence, accidental symmetry operations induced by
the Runge-Lenz vector in general lead out of the domain of the Hamiltonian, and
thus no longer represent physical symmetries. For rational deficit angles, on
the other hand, repeated applications of the Runge-Lenz vector eventually lead
back into the domain of the Hamiltonian and thus give rise to remnant
degeneracies. In \cite{AlH09} we have changed the geometry of the Landau level
problem by confining the charged particle to a 2-d spatial torus, threaded by
$n_\Phi$ units of quantized magnetic flux. The spatial boundary conditions are
then characterized by two self-adjoint extension parameters (related to the two
holonomies of the torus) which explicitly break translation invariance to a
discrete $\Z(n_\Phi)$ symmetry. The continuous accidental symmetry of the
problem in the infinite volume is then reduced to a finite magnetic translation
group and the infinite degeneracy of the corresponding Landau levels is reduced
to $n_\Phi$.

Spatial boundary conditions are naturally characterized by families of
self-adjoint extension parameters \cite{Neu32,Ree75}. For example, the general
Robin boundary conditions of a non-relativistic quantum mechanical particle in
a box with perfectly reflecting walls are characterized by a real-valued
self-adjoint extension parameter $\gamma$, which interpolates continuously
between Dirichlet and Neumann boundary conditions \cite{Bal70,Car90,Bon01}. In
\cite{AlH12} we have used the theory of self-adjoint extensions to derive a
generalized Heisenberg uncertainty relation for finite-volume quantum dots with
general boundary conditions. Perfectly reflecting boundary conditions for
relativistic fermions described by the Dirac equation have also been
investigated in \cite{AlH12}. They are characterized by a 4-parameter family of
self-adjoint extensions. The same is true for the Pauli equation, that arises
from the Dirac equation in the non-relativistic limit. In \cite{AlH13a} we have
investigated the supersymmetric descendants of self-adjointly extended quantum
mechanical Hamiltonians. In particular, we found that the superpartners of
Hamiltonians obeying general Robin boundary conditions characterized by the
parameter $\gamma$ always obey simple Dirichlet boundary conditions, while
$\gamma$ determines the value of the superpotential at the boundary.

Hydrogen atoms confined to a finite volume have been investigated to mimic
the effects of high pressure, in order to better understand, for example, white
dwarf stars. In order to model hydrogen at high pressure, a hydrogen atom at
the center of a spherical cavity has been studied by Michels, de Boer, and Bijl
as early as 1937 \cite{Mic37}. This work was extended by Sommerfeld and Welker
in 1938 \cite{Som38} as well as in \cite{Gro46,Wig54,Fow84,Fro87}. The effect
of general Robin boundary conditions (characterized by the real-valued
self-adjoint extension parameter $\gamma$) on the accidental symmetry of the
non-relativistic hydrogen atom and the harmonic oscillator confined to a
spherical cavity have been investigated in \cite{AlH12a,AlH13}. In both cases,
in general the accidental symmetry is lifted. Again, this is because an
application of the Runge-Lenz vector leads out of the domain of the Hamiltonian.
However, as for the corresponding systems on the cone, repeated application of
the Runge-Lenz vector leads back into the domain of the Hamiltonian, which
gives rise to a finite-volume remnant of the accidental symmetry, at least for
particular radii of the spherical cavity and for specific values of the
self-adjoint extension parameter \cite{Pup98,Sch00,Pup02,AlH12a}.

In this paper, we extend this analysis to the relativistic hydrogen atom
described by the Dirac equation
\cite{Gor28,Dar28,Pid29,Bie62,Won82,Coh93,Vil96}.
The accidental $SO(4)$ symmetry of the non-relativistic problem is then
substantially reduced, but not lifted completely. This is due to the
Johnson-Lippman operator \cite{Joh50,Bie62,Bie83}, the relativistic
counter part of the Runge-Lenz vector, which still leads to enhanced accidental
degeneracies. When the relativistic hydrogen atom is placed inside a confining
spherical cavity, the accidental symmetry is completely
lifted, because an application of the Johnson-Lippman operator again leads
outside the domain of the Hamiltonian. In contrast to the non-relativistic case,
repeated applications of the Johnson-Lippman operator do not lead back into the
domain of the Hamiltonian, and thus no remnant accidental symmetry persists in
a finite volume. The non-relativistic limit of the Dirac hydrogen atom leads to
the Pauli equation (and not to the Schr\"odinger equation without spin).
Interestingly, the conserved current induced by the Dirac equation contains a
spin contribution. This again gives rise to self-adjoint extensions, which in
general induce spin-orbit couplings via the spatial boundary conditions.
Remarkably, in this case again a finite-volume remnant of the accidental
symmetry persists for particular radii of the confining cavity and for
particular values of the self-adjoint extension parameter. Our study further
illuminates the subtle effects of self-adjoint extension parameters on
accidental symmetries, in particular, for relativistic quantum systems and
their non-relativistic counter parts.

The rest of the paper is organized as follows. Section 2 reviews the accidental
symmetries of the hydrogen atom, both in the non-relativistic Schr\"odinger and
in the relativistic Dirac treatment. In Section 3 we discuss the self-adjoint
extension parameters characterizing perfectly reflecting cavity walls both in
the relativistic case and in the non-relativistic case with and without spin.
In Section 4 we place the hydrogen atom in a spherical cavity. After reviewing
the results of the non-relativistic Schr\"odinger treatment, we proceed to the
Dirac equation and its non-relativistic Pauli equation limit. Finally, Section
5 contains our conclusions.

\section{Accidental Symmetries of the Hydrogen Atom}

In order to make the paper self-contained, in this section we review the
accidental symmetries of both the relativistic and the non-relativistic
hydrogen atom. In the non-relativistic Schr\"odinger problem the accidental
symmetry is generated by the Runge-Lenz vector, while in the relativistic Dirac
problem it is generated by the Johnson-Lippmann operator.

\subsection{Accidental Symmetry of the Non-re\-la\-ti\-vis\-tic Schr\"odinger
Atom}

Let us consider the Schr\"odinger Hamiltonian for the hydrogen atom
\begin{equation}
H = - \frac{1}{2 M} \Delta - \frac{e^2}{r}.
\end{equation}
Here $M$ is the electron mass and $e$ is the fundamental electric charge unit.
Throughout this paper we put $\hbar = 1$ but leave the velocity of light $c$
explicit in order to facilitate considerations of the non-relativistic limit.
Thanks to the isotropy of the Coulomb potential, the Hamiltonian obviously
commutes with the angular momentum $\vec L = \vec r \times \vec p$, i.e.\
$[H,\vec L] = 0$. It is well known, but not so obvious, that the Hamiltonian
also commutes with the Runge-Lenz vector
\begin{equation}
\label{Rungedef}
\vec R = \frac{1}{2 M} \left(\vec p \times \vec L - \vec L \times \vec p\right)
- e^2 \vec e_r, \quad \vec e_r = \frac{\vec r}{r}.
\end{equation}
The angular momentum $\vec L$ and the Runge-Lenz vector $\vec R$
generate an $SO(4)$ extension of the $SO(3)$ rotational symmetry, with the
following commutation relations
\begin{equation}
[L_i,L_j] = i \varepsilon_{ijk} L_k, \quad [L_i,R_j] = i \varepsilon_{ijk} R_k,
\quad [R_i,R_j] = i \frac{2 H}{M} \varepsilon_{ijk} L_k.
\end{equation}
The two Casimir operators of the $SO(4)$ algebra are
\begin{equation}
C_1 = \vec L \, ^2 - \frac{M}{2 H} \vec R \, ^2 = - \frac{M e^4}{2 H}, \quad
C_2 = \vec L \cdot \vec R.
\end{equation}
Using the explicit form of $\vec R$ one can show that $\vec L \cdot \vec R = 0$,
which implies that only particular representations of $SO(4)$ can be realized
in the hydrogen atom. These representations are $n^2$-fold degenerate, which
indeed represents the correct accidental degeneracy of the hydrogen atom bound
states. For example, the 2s state is degenerate with the three 2p states, and
the 3s state is degenerate with the three 3p and the five 3d states. The value
of the Casimir operator in the corresponding representations is $C_1 = n^2$,
such that
\begin{equation}
H = - \frac{M e^4}{2 C_1} \ \Rightarrow \ E = - \frac{M e^4}{2 n^2},
\end{equation}
which indeed is the familiar spectrum of hydrogen atom bound states.

\subsection{Accidental Symmetry of the Relativistic Dirac Atom}

Let us now proceed to the Dirac Hamiltonian
\begin{equation}
H = \vec \alpha \cdot \vec p c + \beta M c^2 - \frac{e^2}{r},
\end{equation}
using the following conventions for the Dirac matrices
\begin{equation}
\vec \alpha = \left(\begin{array}{cc} \0 & \vec \sigma \\ \vec \sigma & \0
\end{array}\right), \quad
\beta = \left(\begin{array}{cc} \1 & \0 \\ \0 & - \1 \end{array}\right), \quad
\vec \Sigma = \left(\begin{array}{cc} \vec \sigma & \0 \\ \0 & \vec \sigma
\end{array}\right), \quad
\gamma_5 = \left(\begin{array}{cc} \0 & \1 \\ \1 & \0 \end{array}\right).
\end{equation}
Here $\1$ and $\0$ represent $2 \times 2$ unit- and zero-matrices, and
$\vec \sigma$ are the Pauli matrices. The Hamiltonian commutes with the total
angular momentum
\begin{equation}
\vec J = \vec L + \vec S = \vec L + \frac{\vec \Sigma}{2}, \quad
[H,\vec J] = 0.
\end{equation}
Furthermore, it is parity invariant and thus it commutes with the operator
\begin{equation}
P = \beta I, \quad [H,P] = 0, \quad [\vec J,P] = 0,
\end{equation}
where $I$ performs a spatial inversion. Dirac has also discovered another
symmetry of the system generated by the operator
\begin{equation}
K = \beta(\vec \Sigma \cdot \vec L + 1), \quad [H,K] = 0, \quad [\vec J,K] = 0,
\quad [P,K] = 0.
\end{equation}
The eigenvalues $k$ of $K$ follow from
\begin{eqnarray}
&&\vec \sigma \cdot \vec L + \1 =
(\vec L + \vec S)^2 - \vec L \, ^2 - \vec S \, ^2 + \1 =
\vec J \, ^2 - \vec L \, ^2 + \frac{\1}{4} \ \Rightarrow \nonumber \\
&&k = j(j+1) - l(l+1) + \frac{1}{4} = j(j+1) - \left(j \mp \frac{1}{2}\right)
\left(j \mp \frac{1}{2} + 1\right) + \frac{1}{4} \nonumber \\
&& \quad = \pm \left(j + \frac{1}{2}\right).
\end{eqnarray}
The sign of the quantum number $k$ indicates whether $\vec L$ and $\vec S$
couple to $j = l + \frac{1}{2}$ ($k = j + \frac{1}{2}$) or to
$j = l - \frac{1}{2}$ ($k = - (j + \frac{1}{2})$). Energy eigenstates are
characterized by the principal quantum number $n$ as well as by $j$, $j_3$, and
$k$, such that
\begin{eqnarray}
&&H|n j j_3 k\rangle = E_{nk} |n j j_3 k\rangle, \,
\vec J \, ^2 |n j j_3 k\rangle = j(j+1) |n j j_3 k\rangle, \,
J_3 |n j j_3 k\rangle = j_3 |n j j_3 k\rangle, \nonumber \\
&&K |n j j_3 k\rangle = k |n j j_3 k\rangle.
\end{eqnarray}
Note that in some other works the eigenvalue of $K$ is denoted as $-k$.
The energy eigenvalues (of positive energy states) are given by
\cite{Gor28,Dar28}
\begin{equation}
E_{nk} = M c^2
\left(1 + \frac{\alpha^2}{(n - |k| + \sqrt{k^2 - \alpha^2})^2}\right)^{-1/2}.
\end{equation}
Here $\alpha = e^2/c$ is the fine-structure constant (in units were
$\hbar = 1$). In the non-relativistic limit one recovers the Schr\"odinger
result, corrected by spin-orbit fine-structure effects
\begin{equation}
E_{nk} = M c^2 \left[1 - \frac{e^4}{2 n^2 c^2} -
\frac{e^8}{2 n^3 c^4}
\left(\frac{1}{j + \frac{1}{2}} - \frac{3}{4n}\right) + \dots \right].
\end{equation}
The fine-structure leads to a partial lifting of the large $n^2$-fold degeneracy
related to the accidental $SO(4)$ symmetry of the non-relativistic problem
(without spin). In particular, the energy no longer just depends on $n$ but also
on $j$ and, for a given $n$, is slightly larger for the states with larger $j$.
Still, a remnant of the accidental symmetry persists even for the Dirac
equation, because the energy depends only on $|k|$. As a result, the states with
$j = l + \frac{1}{2}$ ($k = j + \frac{1}{2}$), which have orbital angular
momentum $l$, and the states with $j = l + 1 - \frac{1}{2}$
($k = - (j + \frac{1}{2})$), which have orbital angular momentum $l+1$, are
degenerate (as long as they have the same principal quantum number $n$). This
gives rise to an accidental $2(2j+1)$-fold degeneracy. The states
with the maximal orbital angular momentum $l = n - 1$ and the
maximum total angular momentum $j = l + \frac{1}{2} = n - \frac{1}{2}$ only have
the usual $(2j+1)$-fold degeneracy that results from the $SO(3)$ rotational
symmetry. For example, the 4 states 2S$_{1/2}$ and 2P$_{1/2}$ with total angular
momentum $j = \frac{1}{2}$ (in the standard spectroscopic $nl_j$ notation) are
degenerate while the 4 states 2P$_{3/2}$ with $j = \frac{3}{2}$ have a slightly
higher energy. Similarly, the 4 states 3S$_{1/2}$ and 3P$_{1/2}$ are accidentally
degenerate, the 8 states 3P$_{3/2}$ and 3D$_{3/2}$ are again accidentally
degenerate, but the 6 states 3D$_{5/2}$ with the maximal
$j = 3 - \frac{1}{2} = \frac{5}{2}$ are not.

The relativistic counterpart of the Runge-Lenz vector, which again commutes with
the Hamiltonian, is the pseudo-scalar Johnson-Lippmann operator
\cite{Joh50,Bie62,Bie83}
\begin{equation}
A = i K \gamma_5 \left(\frac{H}{M c^2} - \beta\right) -
\alpha \vec \Sigma \cdot \vec e_r, \quad [H,A] = 0, \quad [\vec J,A] = 0.
\end{equation}
The operator $A$ anti-commutes with $K$, i.e.\ $\{K,A\} = 0$, while $A^2$,
which plays the role of a supersymmetric ``Hamiltonian''
\cite{Suk85,Jar86,Dah95,Kir03} is directly related to $H$ and $K$ via
\begin{equation}
\label{susyH}
A^2 = K^2 \left[\left(\frac{H}{M c^2}\right)^2 - 1\right] + \alpha^2.
\end{equation}
The eigenvalues of $A^2$ are hence given by
\begin{equation}
{\rm a}^2 = k^2 \left[\left(\frac{E_{nk}}{M c^2}\right)^2 - 1\right] + \alpha^2
= \alpha^2 - \frac{\alpha^2 k^2}{(n - |k| + \sqrt{k^2 - \alpha^2})^2 + \alpha^2}.
\end{equation}
The operator $A$ acts on energy eigenstates as
\begin{equation}
A|n j j_3 k\rangle = {\rm a} |n j j_3 -k\rangle,
\end{equation}
i.e.\ it relates two accidentally degenerate states with quantum numbers
$\pm k$. If a state has maximal $j = n - \frac{1}{2}$, it is annihilated by the
operator $A$ because then $n = j + \frac{1}{2} = k = |k|$ such that
\begin{equation}
{\rm a}^2 = \alpha^2 - \frac{\alpha^2 k^2}{k^2 - \alpha^2 + \alpha^2} = 0 \
\Rightarrow \ {\rm a} = 0.
\end{equation}
As was pointed out in \cite{Dah95}, the Johnson-Lippman operator generates
a supersymmetry that relates the accidentally degenerate states. The unpaired
state with maximal $j = n - \frac{1}{2}$ is the lowest state underneath a tower
of paired states with the same value of $j$ but higher values of $n$. The fact
that the lowest state is not paired indicates that supersymmetry is not
spontaneously broken.

\section{Perfectly Reflecting Cavity Boundary Conditions}

Following \cite{AlH12}, in this section we discuss perfectly reflecting boundary
conditions, first in the context of the non-relativistic Schr\"odinger equation
and then for the relativistic Dirac equation. Next we take the non-relativistic
limit of the Dirac equation and arrive at the Pauli equation. The spin then
enters the conserved current, with important implications for the self-adjoint
extension parameters that characterize the spatial boundary conditions.

\subsection{Non-relativistic Robin Boundary Conditions}

Let us consider the general Schr\"odinger Hamiltonian
\begin{equation}
H = \frac{{\vec p \,}^2}{2 M} + V(\vec x) = - \frac{1}{2 M} \Delta + V(\vec x).
\end{equation}
The continuity equation that guarantees probability conservation then takes the
form
\begin{equation}
\p_t \rho(\vec x,t) + \vec \nabla \cdot \vec j(\vec x,t) = 0,
\end{equation}
where the probability density and the corresponding current density are given by
\begin{equation}
\rho(\vec x,t) = |\Psi(\vec x,t)|^2, \quad
\vec j(\vec x,t) = \frac{1}{2 M i}
\left[\Psi(\vec x,t)^* \vec \nabla \Psi(\vec x,t) -
\vec \nabla \Psi(\vec x,t)^* \Psi(\vec x,t)\right].
\end{equation}

We now consider an arbitrarily shaped spatial region $\Omega$ and we demand that
no probability leaks outside this region. This is ensured when the component of
the probability current normal to the surface vanishes
\begin{equation}
\vec n(\vec x) \cdot \vec j(\vec x) = 0, \quad \vec x \in \p \Omega.
\end{equation}
Here $\vec n(\vec x)$ is the unit-vector normal to the surface $\p \Omega$ at
the point $\vec x$. The most general so-called Robin boundary condition that
ensures probability conservation is given by
\begin{equation}
\label{bcdot}
\gamma(\vec x) \Psi(\vec x) + \vec n(\vec x) \cdot \vec \nabla \Psi(\vec x) = 0,
\quad x \in \p \Omega,
\end{equation}
which indeed implies
\begin{equation}
\vec n(\vec x) \cdot \vec j(\vec x) = \frac{1}{2 M i}
[- \Psi(\vec x)^* \gamma(\vec x) \Psi(\vec x,t) +
\gamma(\vec x)^* \Psi(\vec x,t)^* \Psi(\vec x,t)] = 0, \quad
\vec x \in \p \Omega,
\end{equation}
such that $\gamma(\vec x) \in \R$.

It is important to show that the Hamiltonian endowed with the boundary
condition eq.(\ref{bcdot}) is indeed self-adjoint. First, we investigate
\begin{eqnarray}
\langle\chi|H|\Psi\rangle&=&
\int_\Omega d^3x \ \chi(\vec x)^*
\left[- \frac{1}{2 M} \Delta + V(\vec x)\right] \Psi(\vec x) \nonumber \\
&=&\int_\Omega d^3x \
\left[\frac{1}{2 M} \vec \nabla \chi(\vec x)^* \cdot \vec \nabla \Psi(\vec x) +
\chi(\vec x)^* V(\vec x) \Psi(\vec x) \right] \nonumber \\
&-&\frac{1}{2 M}
\int_{\p \Omega} d\vec n \cdot \chi(\vec x)^* \vec \nabla \Psi(\vec x)
\nonumber \\
&=&\int_\Omega d^3x \ \left\{\left[- \frac{1}{2 M} \Delta + V(\vec x)\right]
\chi(\vec x)^* \right\} \Psi(\vec x) \nonumber \\
&+&\frac{1}{2 M} \int_{\p \Omega} d\vec n \cdot
\left[\vec \nabla \chi(\vec x)^* \Psi(\vec x) -
\chi(\vec x)^* \vec \nabla \Psi(\vec x)\right] \nonumber \\
&=&\langle\Psi|H|\chi\rangle^* + \frac{1}{2 M} \int_{\p \Omega} d\vec n \cdot
\left[\vec \nabla \chi(\vec x)^* \Psi(\vec x) -
\chi(\vec x)^* \vec \nabla \Psi(\vec x)\right].
\end{eqnarray}
Consequently, the Hamiltonian is Hermitean (or symmetric in mathematical
parlance) if
\begin{equation}
\label{symmetricHdot}
\int_{\p \Omega} d\vec n \cdot
\left[\vec \nabla \chi(\vec x)^* \Psi(\vec x) -
\chi(\vec x)^* \vec \nabla \Psi(\vec x)\right] = 0.
\end{equation}
Using the boundary condition eq.(\ref{bcdot}), eq.(\ref{symmetricHdot})
simplifies to
\begin{equation}
\int_{\p \Omega} d^2x \left[\vec n(\vec x) \cdot \vec \nabla \chi(\vec x)^* +
\gamma(\vec x) \chi(\vec x)^*\right] \Psi(\vec x) = 0.
\end{equation}
$\Psi(\vec x)$ itself is not restricted at the boundary, and hence the
Hermiticity of $H$ requires
\begin{equation}
\vec n(\vec x) \cdot \vec \nabla \chi(\vec x) + \gamma(\vec x)^* \chi(\vec x) =
0.
\end{equation}
For $\gamma(\vec x) \in \R$, this is again the boundary condition of
eq.(\ref{bcdot}). Since $\chi(\vec x)$ must obey the same boundary condition as
$\Psi(\vec x)$, the domain of $H^\dagger$, $D(H^\dagger)$, coincides with
the domain of $H$, $D(H)$. Since $D(H^\dagger) = D(H)$, the Hamiltonian is not
only Hermitean but, in fact, self-adjoint.

\subsection{Relativistic Cavity Boundary Conditions}

A Dirac particle in a 1-d box has been considered in \cite{Alb96,Alb11}, and
general boundary conditions have been investigated in \cite{Alo97}. In
\cite{AlH12} we have investigated general perfectly reflecting boundary
conditions for Dirac fermions in a 3-d spatial region $\Omega$, thereby
generalizing the standard MIT bag boundary conditions \cite{Cho74,Cho74a,Has78}.

To keep the discussion as general as possible, we investigate Dirac fermions
coupled to an external static electromagnetic field, such that the Hamiltonian
is given by
\begin{equation}
H = \vec \alpha \cdot \left(\vec p c + e \vec A(\vec x)\right) + \beta M c^2 -
e \Phi(\vec x) = - i \vec \alpha \cdot \vec D c + \beta M c^2 - e \Phi(\vec x).
\end{equation}
Here $\Phi(\vec x)$ is the scalar and $\vec A(\vec x)$ is the vector potential.
The covariant derivative then takes the form
\begin{equation}
\vec D = \vec \nabla + i \frac{e}{c} \vec A(\vec x).
\end{equation}
The Hamiltonian acts on a 4-component Dirac spinor $\Psi(\vec x,t)$. In
the relativistic case, the continuity equation is given by
\begin{equation}
\p_t \rho(\vec x,t) + \vec \nabla \cdot \vec j(\vec x,t) = 0,
\end{equation}
where
\begin{equation}
\rho(\vec x,t) = \Psi(\vec x,t)^\dagger \Psi(\vec x,t), \quad
\vec j(\vec x,t) = c \Psi(\vec x,t)^\dagger \vec \alpha \Psi(\vec x,t).
\end{equation}
Under time-independent gauge transformations, the gauge and fermion fields
transform as
\begin{equation}
^\varphi\Phi(\vec x) = \Phi(\vec x), \
^\varphi\vec A(\vec x) = \vec A(\vec x) - \vec \nabla \varphi(\vec x), \
^\varphi\Psi(\vec x) = \exp\left(i \frac{e}{c} \varphi(\vec x)\right)
\Psi(\vec x).
\end{equation}

Just as in the non-relativistic case, we investigate the Hermiticity of the
Hamiltonian by considering
\begin{eqnarray}
\langle \chi|H|\Psi\rangle&=&\int_\Omega d^3x \ \chi(\vec x)^\dagger
\left[\vec \alpha \cdot \left(- i c \vec \nabla + e \vec A(\vec x)\right) +
\beta M c^2 - e \Phi(\vec x)\right] \Psi(\vec x) \nonumber \\
&=&\int_\Omega d^3x \ \left\{\left[\vec \alpha \cdot
\left(- i c \vec \nabla + e \vec A(\vec x)\right) + \beta M c^2 - e \Phi(\vec x)
\right]\chi(\vec x)\right\}^\dagger \Psi(\vec x) \nonumber \\
&-&i c \int_{\p \Omega} d\vec n \cdot \chi(\vec x)^\dagger \vec \alpha
\Psi(\vec x) \nonumber \\
&=&\langle \Psi|H|\chi\rangle^* -
i c \int_{\p \Omega} d\vec n \cdot \chi(\vec x)^\dagger \vec \alpha \Psi(\vec x).
\end{eqnarray}
This implies the Hermiticity condition
\begin{equation}
\label{Dirachermiticity}
\chi(\vec x)^\dagger \vec n(\vec x) \cdot \vec \alpha \Psi(\vec x) = 0,
\quad \vec x \in \p \Omega.
\end{equation}
The corresponding appropriate self-adjoint extension condition is given by
\begin{equation}
\label{Diracselfadjointness}
\left(\begin{array}{c} \Psi_3(\vec x) \\ \Psi_4(\vec x) \end{array}\right) =
\lambda(\vec x)
\left(\begin{array}{c} \Psi_1(\vec x) \\ \Psi_2(\vec x) \end{array}\right),
\quad \lambda(\vec x) \in GL(2,\C), \quad \vec x \in \p \Omega,
\end{equation}
which reduces eq.(\ref{Dirachermiticity}) to
\begin{eqnarray}
&&\chi(\vec x)^\dagger
\left(\begin{array}{cc} \0 & \vec n(\vec x) \cdot \vec \sigma \\
\vec n(\vec x) \cdot \vec \sigma & \0 \end{array}\right) \Psi(\vec x) =
\nonumber \\
&&\left[\left(\chi_1(\vec x)^*,\chi_2(\vec x)^* \right)
\vec n(\vec x) \cdot \vec \sigma \lambda(\vec x) +
\left(\chi_3(\vec x)^*,\chi_4(\vec x)^* \right)
\vec n(\vec x) \cdot \vec \sigma \right]
\left(\begin{array}{c} \Psi_1(\vec x) \\ \Psi_2(\vec x) \end{array}\right)
= 0 \ \Rightarrow \nonumber \\
&&\left(\begin{array}{c} \chi_3(\vec x) \\ \chi_4(\vec x) \end{array}\right) =
- \vec n(\vec x) \cdot \vec \sigma \lambda(\vec x)^\dagger
\vec n(\vec x) \cdot \vec \sigma
\left(\begin{array}{c} \chi_1(\vec x) \\ \chi_2(\vec x) \end{array}\right),
\end{eqnarray}
Self-adjointness of $H$ requires that $D(H) = D(H^\dagger)$, which demands
\begin{equation}
\lambda(\vec x) = - \vec n(\vec x) \cdot \vec \sigma \lambda(\vec x)^\dagger
\vec n(\vec x) \cdot \vec \sigma \ \Rightarrow \
\vec n(\vec x) \cdot \vec \sigma \lambda(\vec x) = -
\left[\vec n(\vec x) \cdot \vec \sigma \lambda(\vec x)\right]^\dagger.
\end{equation}
Consequently, $\vec n(\vec x) \cdot \vec \sigma \lambda(\vec x)$ must be
anti-Hermitean. This means that, in contrast to the non-relativistic
Schr\"odinger problem, relativistic Dirac fermions have a 4-parameter family of
self-adjoint extensions characterizing a perfectly reflecting wall. In the
standard MIT bag model \cite{Cho74,Cho74a,Has78} the boundary condition
corresponds to $\lambda(\vec x) = i \vec n(\vec x) \cdot \vec \sigma$. This is
a natural choice, because it maintains spatial rotation invariance around the
normal $\vec n(\vec x)$ on the boundary, but it is not the most general
possibility. As it should be, the self-adjointness condition
eq.(\ref{Diracselfadjointness}) is gauge covariant and it ensures that
\begin{eqnarray}
\vec n(\vec x) \cdot \vec j(\vec x)\!\!\!\!&=&\!\!\!\!c \Psi(\vec x)^\dagger
\left(\begin{array}{cc} \0 & \vec n(\vec x) \cdot \vec \sigma \\
\vec n(\vec x) \cdot \vec \sigma & \0 \end{array}\right) \Psi(\vec x) =
\nonumber \\
&=&\!\!\!\!c \left[\left(\Psi_1(\vec x)^*,\Psi_2(\vec x)^* \right)
\vec n(\vec x) \cdot \vec \sigma \lambda(\vec x) +
\left(\Psi_3(\vec x)^*,\Psi_4(\vec x)^* \right)
\vec n(\vec x) \cdot \vec \sigma \right]
\left(\begin{array}{c} \Psi_1(\vec x) \\ \Psi_2(\vec x) \end{array}\right)
\nonumber \\
&=&\!\!\!\!c \left(\Psi_1(\vec x)^*,\Psi_2(\vec x)^* \right) \left[
\vec n(\vec x) \cdot \vec \sigma \lambda(\vec x) +
\lambda(\vec x)^\dagger \vec n(\vec x) \cdot \vec \sigma \right]
\left(\begin{array}{c} \Psi_1(\vec x) \\ \Psi_2(\vec x) \end{array}\right) = 0.
\end{eqnarray}

\subsection{Non-relativistic Boundary Conditions with Spin}

Following \cite{AlH12}, we now consider the non-relativistic limit, in which
the lower components of the Dirac spinor reduce to
\begin{equation}
\label{nrlimit}
\left(\begin{array}{c} \Psi_3(\vec x) \\ \Psi_4(\vec x) \end{array} \right) =
\frac{\vec \sigma \cdot \left(\vec p c + e \vec A(\vec x)\right)}{2 M c^2}
\left(\begin{array}{c} \Psi_1(\vec x) \\ \Psi_2(\vec x) \end{array} \right) =
\frac{1}{2 M c \, i} \vec \sigma \cdot \vec D \Psi(\vec x).
\end{equation}
Here the 2-component Pauli spinor is given by
\begin{equation}
\Psi(\vec x) =
\left(\begin{array}{c} \Psi_1(\vec x) \\ \Psi_2(\vec x) \end{array} \right).
\end{equation}
To leading order, the Dirac Hamiltonian then reduces to the Pauli Hamiltonian
\begin{equation}
\label{PauliH}
H = M c^2 + \frac{\left(\vec p c + e \vec A(\vec x)\right)^2}{2 M c^2} -
e \Phi(\vec x) + \mu \vec \sigma \cdot \vec B(\vec x),
\end{equation}
with $\vec B(\vec x) = \vec \nabla \times \vec A(\vec x)$ being the magnetic
field and $\mu = e/2 M c$ being the Bohr magneton, i.e.\ the magnetic moment
of the electron. Here we neglect higher order contributions such as spin-orbit
couplings and the Darwin term.

The self-adjoint extension parameters that characterize the most general
perfectly reflecting boundary condition crucially depend on the form of the
conserved current. Therefore we now use eq.(\ref{nrlimit}) to obtain
\begin{eqnarray}
\vec j(\vec x)&=&c
\left(\Psi_1(\vec x)^*,\Psi_2(\vec x)^*,\Psi_3(\vec x)^*,\Psi_4(\vec x)^*\right)
\left(\begin{array}{cc} \0 & \vec \sigma \\ \vec \sigma & \0 \end{array}\right)
\left(\begin{array}{c} \Psi_1(\vec x) \\ \Psi_2(\vec x) \\
\Psi_3(\vec x) \\ \Psi_4(\vec x) \end{array} \right) \nonumber \\
&=&\frac{1}{2 M i} \left[\Psi(\vec x)^\dagger \vec D \Psi(\vec x) -
(\vec D \Psi(\vec x))^\dagger \Psi(\vec x)\right] \nonumber \\
&-&\frac{1}{2 M}
\left[\Psi(\vec x,t)^\dagger \vec \sigma \times \vec D \Psi(\vec x,t) -
(\vec D \Psi(\vec x,t))^\dagger \times \vec \sigma \Psi(\vec x,t)\right]
\nonumber \\
&=&\frac{1}{2 M i}
\left[\Psi(\vec x,t)^\dagger \vec D \Psi(\vec x,t) -
(\vec D \Psi(\vec x,t))^\dagger \Psi(\vec x,t)\right] \nonumber \\
&+&\frac{1}{2 M} \vec \nabla \times
\left[\Psi(\vec x,t)^\dagger \vec \sigma \Psi(\vec x,t)\right].
\end{eqnarray}
Besides the non-relativistic probability current that is familiar from the
Schr\"odinger equation, in the context of the Pauli equation an additional
spin contribution arises from the Dirac current. Since it is a curl, the spin
contribution is automatically divergenceless, and the continuity equation is
given by
\begin{equation}
\p_t \rho(\vec x,t) + \vec \nabla \cdot \vec j(\vec x,t) = 0,
\end{equation}
with the probability density
$\rho(\vec x,t) = \Psi(\vec x,t)^\dagger \Psi(\vec x,t)$. Although the current
would also be conserved without the spin term, this contribution naturally
belongs to the current that emerges from the Dirac equation.

Again following \cite{AlH12}, we now consider the gauge covariant boundary
condition
\begin{equation}
\label{sadjnonrel}
\gamma(\vec x) \Psi(\vec x) + \vec n(\vec x) \cdot
\left[\vec D  \Psi(\vec x) - i \vec \sigma \times \vec D  \Psi(\vec x)\right] =
0, \quad \gamma(\vec x) \in GL(2,\C), \quad \vec x \in \p \Omega,
\end{equation}
which implies
\begin{eqnarray}
\vec n(\vec x) \cdot \vec j(\vec x)&=&\frac{1}{2 M i}
\left[\Psi(\vec x)^\dagger \vec n(\vec x) \cdot \vec D \Psi(\vec x) -
(\vec n(\vec x) \cdot \vec D \Psi(\vec x))^\dagger \Psi(\vec x)\right]
\nonumber \\
&-&\frac{1}{2 M} \left[\Psi(\vec x,t)^\dagger \vec n \cdot
\left(\vec \sigma \times \vec D \Psi(\vec x,t)\right) -
\vec n \cdot \left((\vec D \Psi(\vec x,t))^\dagger \times \vec \sigma\right)
\Psi(\vec x,t)\right] \nonumber \\
&=&\frac{1}{2 M i}
\left[- \Psi(\vec x)^\dagger \gamma(\vec x) \Psi(\vec x) +
\Psi(\vec x)^\dagger \gamma(\vec x)^\dagger \Psi(\vec x)\right] = 0,
\end{eqnarray}
such that
\begin{equation}
\gamma(\vec x)^\dagger = \gamma(\vec x).
\end{equation}
As in the fully relativistic Dirac problem, also in the non-relativistic Pauli
problem with spin we again obtain a 4-parameter family of self-adjoint
extensions, now parametrized by the $2 \times 2$ Hermitean matrix
$\gamma(\vec x)$.

As in the Schr\"odinger and Dirac cases, partial integration leads to the
Hermiticity condition for the Pauli Hamiltonian of eq.(\ref{PauliH})
\begin{equation}
\label{Hnonrel}
\int_{\p \Omega}  d\vec n \cdot \left[\left(\vec D \chi(\vec x)\right)^\dagger
\Psi(\vec x) - \chi(\vec x)^\dagger \vec D \Psi(\vec x)\right] = 0,
\end{equation}
and one obtains
\begin{equation}
\left(\vec D \chi(\vec x)\right)^\dagger \times \vec \sigma \Psi(\vec x) -
\chi(\vec x)^\dagger \vec \sigma \times \vec D \Psi(\vec x) =
\vec \nabla \times \left(\chi(\vec x)^\dagger \vec \sigma \Psi(\vec x)\right).
\end{equation}
Applying Stoke's theorem and using the fact that the boundary of a boundary is
an empty set (i.e.\ $\p(\p \Omega) = \emptyset$) we arrive at
\begin{eqnarray}
&&\int_{\p \Omega}  d\vec n \cdot
\left[\left(\vec D \chi(\vec x)\right)^\dagger \times \vec \sigma \Psi(\vec x) -
\chi(\vec x)^\dagger \vec \sigma \times \vec D \Psi(\vec x)\right] = \nonumber \\
&&\int_{\p \Omega}  d\vec n \cdot
\vec \nabla \times \left(\chi(\vec x)^\dagger \vec \sigma \Psi(\vec x)\right) =
\int_{\p(\p \Omega)} d\vec l \cdot \chi(\vec x)^\dagger \vec \sigma \Psi(\vec x) =
0.
\end{eqnarray}
The Hermiticity condition eq.(\ref{Hnonrel}) can thus be expressed as
\begin{equation}
\int_{\p \Omega}  d\vec n \cdot \left[\left(\vec D \chi(\vec x) -
i \vec \sigma \times \vec D \chi(\vec x)\right)^\dagger \Psi(\vec x) -
\chi(\vec x)^\dagger \left(\vec D \Psi(\vec x) - i \vec \sigma \times
\vec D \Psi(\vec x) \right)\right] = 0.
\end{equation}
Using the self-adjointness condition eq.(\ref{sadjnonrel}), we then obtain
\begin{equation}
\int_{\p \Omega}  d^2x \ \left[\chi(\vec x)^\dagger \gamma(\vec x)^\dagger
\Psi(\vec x) - \chi(\vec x)^\dagger \gamma(\vec x) \Psi(\vec x) \right] = 0.
\end{equation}
Here we have used the fact that $\gamma(\vec x)$ is Hermitean.

The matrix $\gamma(\vec x)$ results from the matrix $\lambda(\vec x)$ in the
non-re\-la\-ti\-vis\-tic limit. Rewriting the self-adjointness condition
eq.(\ref{sadjnonrel}) as
\begin{equation}
\gamma(\vec x) \Psi(\vec x) + \vec n(\vec x) \cdot \vec \sigma \,
\vec \sigma \cdot \vec D \Psi(\vec x) =
\gamma(\vec x) \Psi(\vec x) + 2 M c \, i \vec n(\vec x) \cdot \vec \sigma
\lambda(\vec x) \Psi(\vec x) = 0,
\end{equation}
one arrives at
\begin{equation}
\gamma(\vec x) = - 2 M c \, i \, \vec n(\vec x) \cdot \vec \sigma
\lambda(\vec x).
\end{equation}
Using the anti-Hermiticity of
$\vec n(\vec x) \cdot \vec \sigma \lambda(\vec x)$, one concludes that
$\gamma(\vec x)$ is indeed Hermitean.

\section{The Hydrogen Atom in a Spherical Cavity}

In this section we place a hydrogen atom at the center of a spherical cavity and
investigate the effect of the boundary conditions on the accidental symmetry.
Again, we first study the Schr\"odinger equation before we proceed to the Dirac
and Pauli equations.

\subsection{The Non-relativistic Schr\"odinger Atom in a Spherical Cavity}

\begin{figure}[tbh]
\begin{center}
\epsfig{file=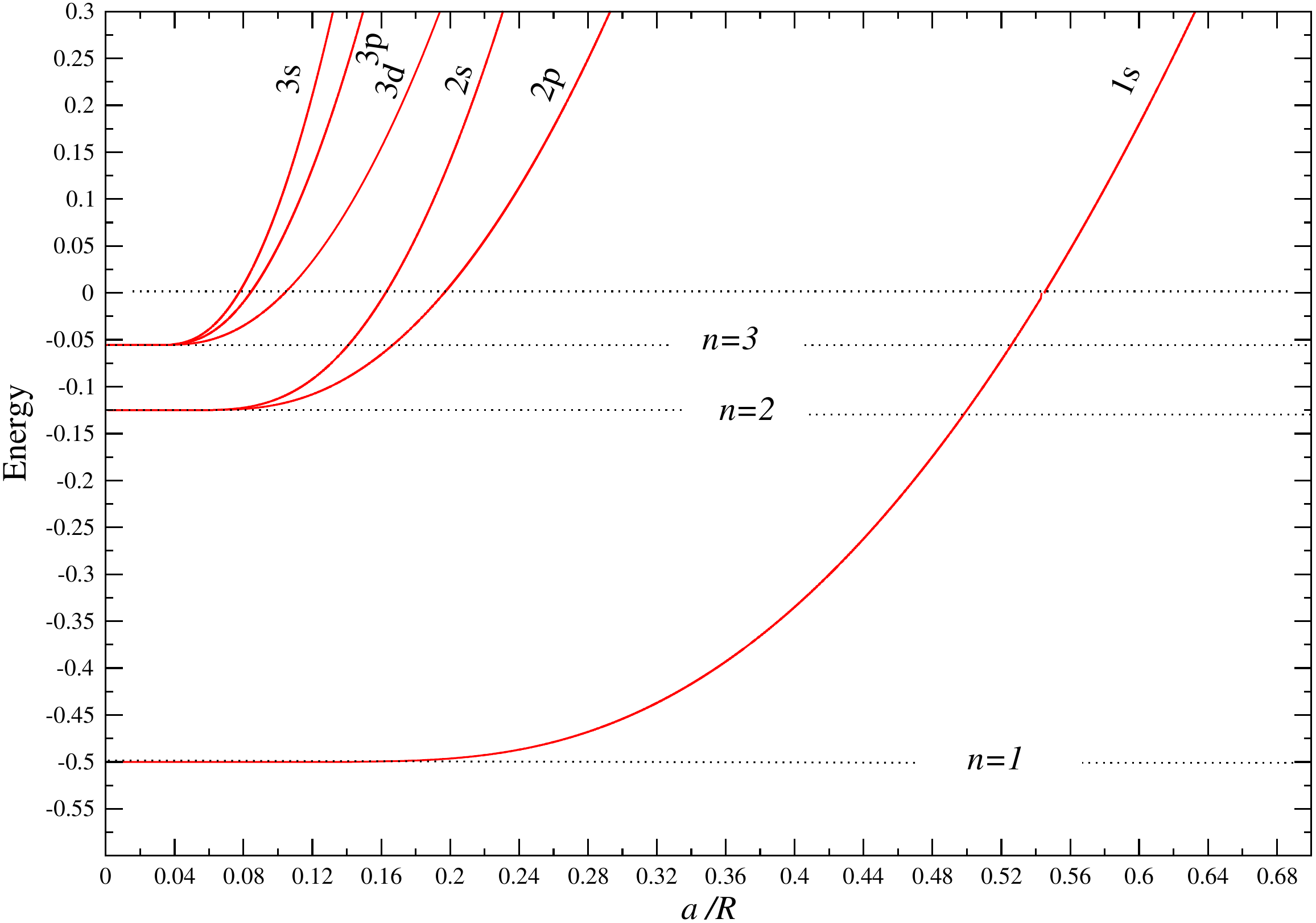,width=10cm}
\end{center}
\caption{\it Spectrum of the Schr\"odinger hydrogen atom centered in a spherical
cavity with the standard Dirichlet boundary condition (i.e.\ $\gamma = \infty$)
as a function of $a/R$. The energies of states with quantum numbers
$n = 1, 2, 3$ are given in units of $M e^4$. The dotted lines represent the
spectrum of the infinite system (cf.\ \cite{AlH12a}).}
\label{specinf}
\end{figure}

\begin{figure}[tbh]
\begin{center}
\epsfig{file=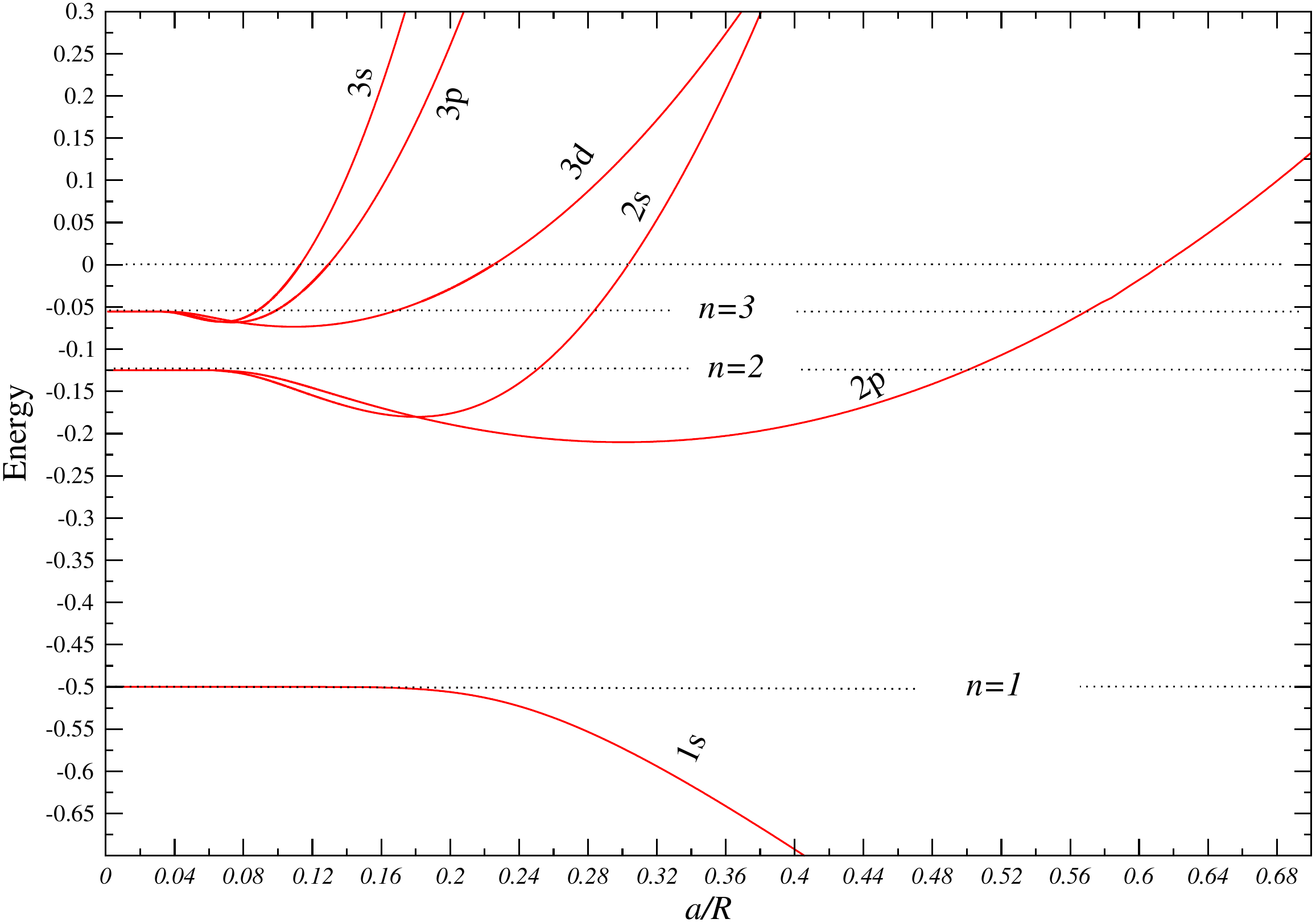,width=10cm}
\end{center}
\caption{\it Spectrum of the Schr\"odinger hydrogen atom centered in a spherical
cavity with Neumann boundary condition (i.e.\ $\gamma = 0$) as a function of
$a/R$. The energies of states with quantum numbers $n = 1, 2, 3$ are given in
units of $M e^4$. The dotted lines represent the spectrum of the infinite
system (c.f.\ \cite{AlH12a}).}
\label{spec0}
\end{figure}

Let us consider the Schr\"odinger Hamiltonian for the hydrogen atom,
\begin{equation}
H = - \frac{1}{2 M} \Delta - \frac{e^2}{r} =
- \frac{1}{2 M} \left(\p_r^2 + \frac{2}{r} \p_r
- \frac{\vec L \, ^2}{r^2}\right) - \frac{e^2}{r}.
\end{equation}
For the wave function we make the factorization ansatz
\begin{equation}
\Psi(\vec x) = \psi_{nl}(r) Y_{lm}(\theta,\varphi),
\end{equation}
which leads to the radial equation
\begin{equation}
\label{radialSE}
\left[- \frac{1}{2 M} \left(\p_r^2 + \frac{2}{r} \p_r
- \frac{l(l + 1)}{r^2}\right) - \frac{e^2}{r}\right] \psi_{nl}(r) =
E \psi_{nl}(r).
\end{equation}
When we write the energy as
\begin{equation}
E = - \frac{M e^4}{2 n^2},
\end{equation}
for the bound state spectrum in the infinite volume $n$ is quantized in
integer units. Inside a spherical cavity, on the other hand, $n$ can take
arbitrary values, and one obtains
\begin{equation}
\psi_{nl}(r) = A \left(\frac{2 r}{n a}\right)^l
L^{2l+1}_{n-l-1}\left(\frac{2 r}{n a}\right)
\exp\left(- \frac{r}{n a}\right),
\end{equation}
where $L^{2l+1}_{n-l-1}(2r/n a)$ is an associated Laguerre function. Imposing
the most general boundary condition, $\gamma(\vec x) \Psi(\vec x) +
\vec n(\vec x) \cdot \vec \nabla \Psi(\vec x) = 0$, one obtains
\begin{equation}
\gamma \psi_{nl}(R) + \p_r  \psi_{nl}(R) = 0,
\end{equation}
where $R$ is the radius of the spherical cavity. Consequently, the quantized
energies result from the equation
\begin{equation}
\left(\frac{\gamma n a}{2} - \frac{1}{2} + \frac{l n a}{2 R}\right)
L^{2l+1}_{n-l-1}\left(\frac{2 R}{n a}\right) -
L^{2l+2}_{n-l-2}\left(\frac{2 R}{n a}\right) = 0.
\end{equation}
The resulting finite volume spectra for $\gamma = \infty$ (i.e.\ Dirichlet
boundary conditions) and for $\gamma = 0$ (i.e.\ Neumann boundary conditions)
are illustrated in Figs.\ \ref{specinf} and \ref{spec0}, respectively.

As we discussed in Section 2.1, in the infinite volume the non-relativistic
hydrogen atom enjoys an accidental $SO(4)$ symmetry generated by the angular
momentum $\vec L$ together with the Runge-Lenz vector $\vec R$ of
eq.(\ref{Rungedef}), which results in an $n^2$-fold degeneracy of the bound
state spectrum. The spectrum of hydrogen confined to a spherical cavity, on the
other hand, no longer shows the accidental degeneracy. Since the boundary
condition does not violate rotation invariance, $\vec L$ is still conserved, but
$\vec R$ is not. This is because the application of the Runge-Lenz vector leads
outside the domain of the Hamiltonian \cite{AlH12a}, as was already pointed out
in \cite{Pup98,Pup02} for the Dirichlet boundary condition with
$\gamma = \infty$. Relying on rotation invariance, we restrict ourselves to
states $\Psi(\vec r) = \psi_{nl}(r) Y_{ll}(\theta,\varphi)$ with the maximal
$m = l$. On such states the raising operator $R_+ = R_x + i R_y$ acts as
\cite{Pup98}
\begin{eqnarray}
\label{Runge}
R_+ \Psi(\vec r)&=&\left[\frac{l + 1}{M} \p_r \psi_{nl}(r) +
\left(e^2 - \frac{l(l + 1)}{M r}\right) \psi_{nl}(r)\right]
Y_{l+1,l+1}(\theta,\varphi) \nonumber \\
&=&\chi_{n,l+1}(r) Y_{l+1,l+1}(\theta,\varphi).
\end{eqnarray}
After an application of $\vec R$ the wave function remains in the domain of the
Hamiltonian only if the new wave function $\chi_{n,l+1}(r)$ also obeys the
boundary condition
\begin{equation}
\label{bcchi}
\gamma \chi_{n,l+1}(R) + \p_r \chi_{n,l+1}(R) = 0.
\end{equation}
Inserting eq.(\ref{Runge}) into this relation and using the boundary condition
$\gamma \psi_{nl}(R) + \p_r \psi_{nl}(R) = 0$ as well as the radial
Schr\"odinger equation (\ref{radialSE}) for $\psi_{nl}(r)$, one finds
\begin{eqnarray}
\label{Rungeselfad}
&&\gamma \chi_{n,l+1}(R) + \p_r \chi_{n,l+1}(R) = \nonumber \\
&&\frac{l + 1}{M} \left[- \gamma \left(\gamma - \frac{2}{R}\right) +
\frac{l(l + 2)}{R^2} - \frac{2 M e^2}{R} - 2 M E \right] \psi_{nl}(R).
\end{eqnarray}
Since the right-hand side of eq.(\ref{Rungeselfad}) does not vanish independent
of the energy $E$, the new wave function $\chi_{n,l+1}(r)$ that results from
the application of $R_+$ on $\psi_{nl}(r)$ does not obey the boundary
condition eq.(\ref{bcchi}), and hence lies outside the domain of the
Hamiltonian. Consequently, the Runge-Lenz vector $\vec R$ leads out of this
domain, and thus no longer represents an accidental symmetry.

\begin{figure}[tbh]
\begin{center}
\epsfig{file=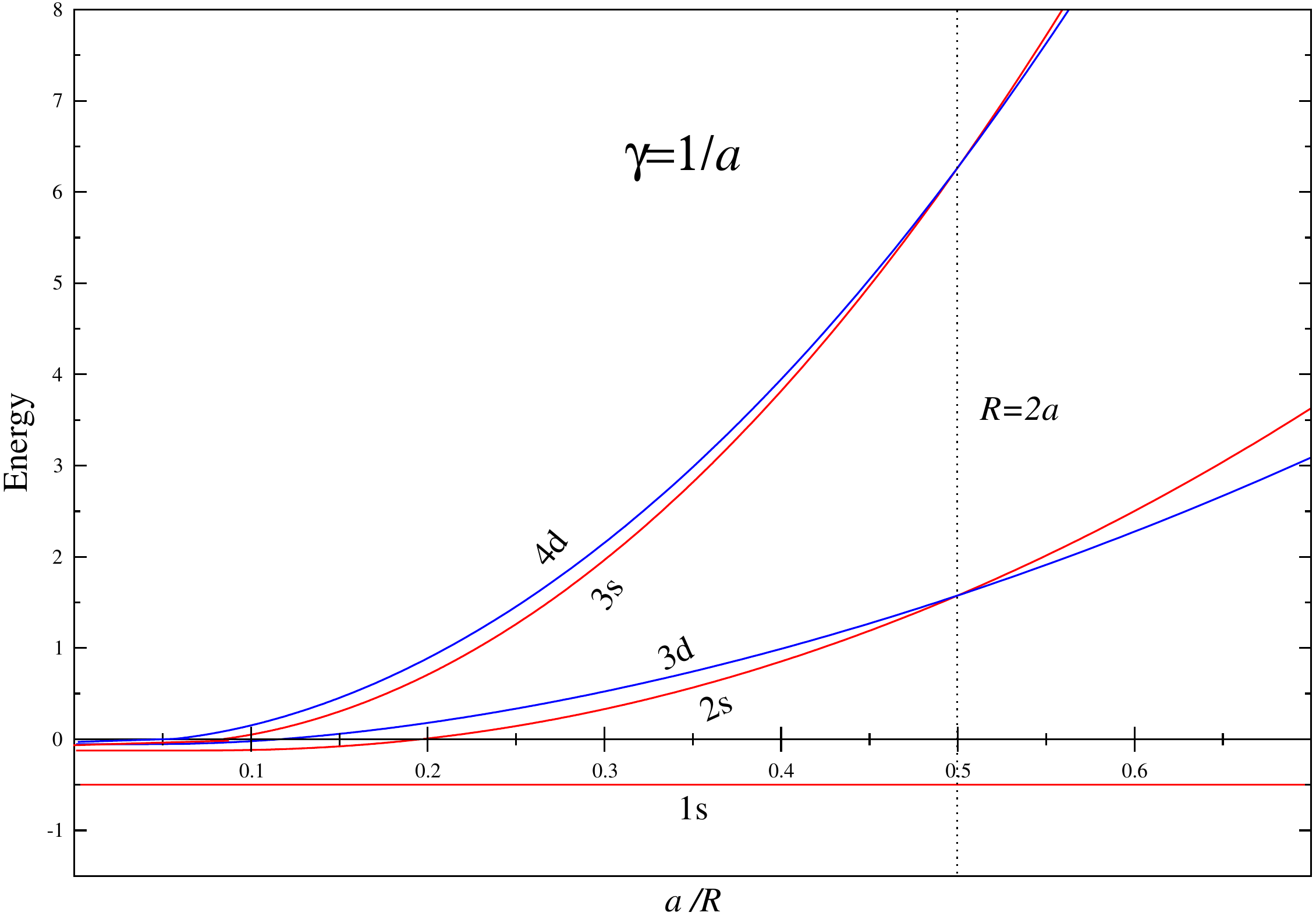,width=10cm} \vskip1.5cm
\epsfig{file=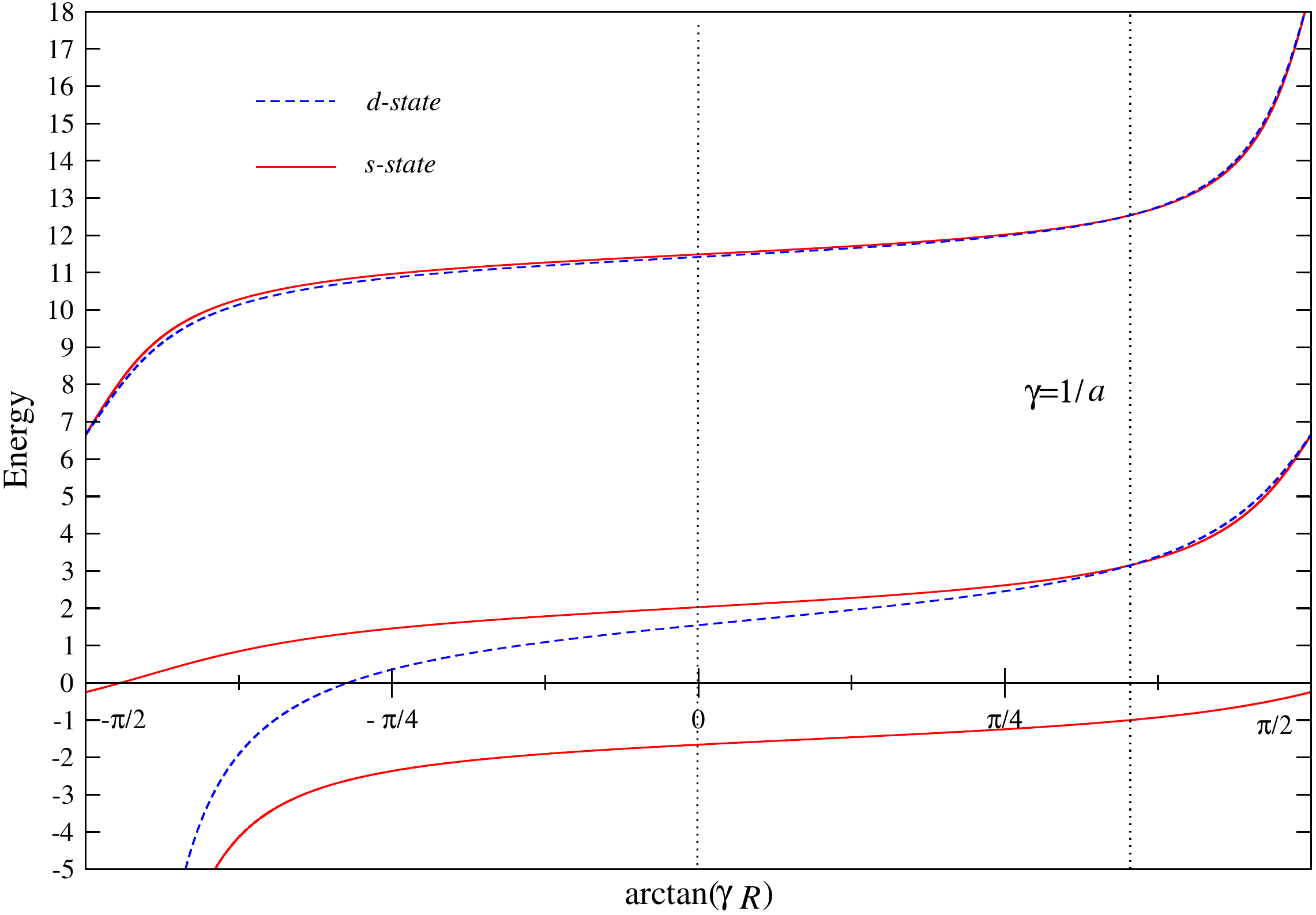,width=10cm}
\end{center}
\caption{\it Top: Energy of s- and d-states for the Schr\"odinger hydrogen atom
in a spherical cavity with $\gamma = 1/a$ as a function of $a/R$. There is an
accidental degeneracy for $R = 2a$. Bottom: Energy of s- and d-states for a
hydrogen atom in a spherical cavity with radius $R = 2a$ as a function of
$\arctan(\gamma R)$. The energies of the 3s and 4d states are very similar for
all values of $\gamma$, but are identical only for $\gamma = 2/R = 1/a$ and for
$\gamma = \pm \infty$. The energy is given in units of $M e^4$. }
\label{accidental}
\end{figure}

Interestingly, for $\gamma = \infty$  a remnant of the accidental symmetry
persists even in a finite volume, however, only for the special cavity radius
$R = (l + 1)(l + 2) a$ \cite{Pup98,Pup02}. Then a state with angular momentum
$l$ is degenerate with a state of angular momentum $l + 2$. In particular, for
$R = 2 a$, the states 2s and 3d, 3s and 4d, 4s and 5d, etc.\ form multiplets of
$1 + 5 = 6$ degenerate states, while for $R = 6 a$, the states 3p and 4f, 4p
and 5f, 5p and 6f, etc.\ form multiplets of $3 + 7 = 10$ degenerate states.
These accidental degeneracies are again due to the Runge-Lenz vector. They
arise because, for the special values $R = (l + 1)(l + 2) a$, the operator
$R_+^2$ maps $\psi_{nl}(r)$ back into the domain of the Hamiltonian. Indeed,
one can show that $R_+^2 \psi_{nl}(r) Y_{ll}(\theta,\varphi) =
\chi_{n,l+2}(r) Y_{l+2,l+2}(\theta,\varphi)$ with
\begin{eqnarray}
\label{chieq}
\chi_{n,l+2}(r)&=&
\frac{2 l + 3}{M}
\left(e^2 - \frac{(l + 1)(l + 2)}{M r}\right) \p_r \psi_{nl}(r) \nonumber \\
&+&\left[\frac{(l+1)(l+2)}{M}
\left(\frac{l(2l + 3)}{M r^2} - \frac{3 e^2}{r} - 2 E\right) +
e^2 \left(e^2 - \frac{l(l + 1)}{Mr}\right)\right] \psi_{nl}(r). \nonumber \\
\,
\end{eqnarray}
Since for $\gamma = \infty$ the wave function obeys the Dirichlet boundary
condition $\psi_{nl}(R) = 0$, for
\begin{equation}
e^2 - \frac{(l + 1)(l + 2)}{M R} = 0 \ \Rightarrow \
R = \frac{(l + 1)(l + 2)}{M e^2} = (l + 1)(l + 2) a,
\end{equation}
the wave function $\chi_{n,l+2}(r)$ indeed obeys the same condition.
Remarkably, the same accidental degeneracy arises for $R = (l + 1)(l + 2) a$ and
$\gamma = 2/R$ \cite{Sch00}. This follows from the fact that then
$\gamma \psi_{nl}(R) + \p_r \psi_{nl}(R) = 0$ indeed implies
$\gamma \chi_{n,l+2}(R) + \p_r \chi_{n,l+2}(R) = 0$. The remnant accidental
degeneracies of the non-relativistic hydrogen atom in a spherical cavity are
illustrated in figure \ref{accidental}.

\subsection{The Relativistic Dirac Atom in a Spherical Cavity}

Let us now consider the Dirac equation in a spherical cavity, i.e.\ we impose
the boundary condition
\begin{equation}
\left(\begin{array}{c} \Psi_3(\vec x) \\ \Psi_4(\vec x) \end{array}\right) =
\lambda(\vec x)
\left(\begin{array}{c} \Psi_1(\vec x) \\ \Psi_2(\vec x) \end{array}\right),
\quad |\vec x| = R,
\end{equation}
where $\vec \sigma \cdot \vec e_r \lambda(\vec x)$ is an anti-Hermitean matrix.
Again, we want to maintain the spherical symmetry, $[H,\vec J] = 0$. We thus
act with $J^a$ on a general Dirac spinor $\Psi(\vec x)$ and demand that
$\Psi(\vec x)' = J^a \Psi(\vec x)$ still obeys the boundary condition
\begin{eqnarray}
\lambda(\vec x)
\left(\begin{array}{c} \Psi_1(\vec x)' \\ \Psi_2(\vec x)' \end{array}\right)&=&
\lambda(\vec x) J^a
\left(\begin{array}{c} \Psi_1(\vec x) \\ \Psi_2(\vec x) \end{array}\right) =
\left(\begin{array}{c} \Psi_3(\vec x)' \\ \Psi_4(\vec x)' \end{array}\right)
\nonumber \\
&=&J^a \left(\begin{array}{c} \Psi_3(\vec x) \\ \Psi_4(\vec x) \end{array}
\right) = J^a \lambda(\vec x)
\left(\begin{array}{c} \Psi_1(\vec x) \\ \Psi_2(\vec x) \end{array} \right).
\end{eqnarray}
This hence implies that $[J^a,\lambda(\vec x)] = 0$. Similarly, the boundary
condition should also maintain the parity symmetry $P$, such that
$\Psi(\vec x)' = P \Psi(\vec x)$ should again obey the boundary condition
\begin{eqnarray}
\lambda(\vec x)
\left(\begin{array}{c} \Psi_1(\vec x)' \\ \Psi_2(\vec x)' \end{array}\right)&=&
\lambda(\vec x)
\left(\begin{array}{c} \Psi_1(- \vec x) \\ \Psi_2(- \vec x) \end{array}\right) =
\left(\begin{array}{c} \Psi_3(\vec x)' \\ \Psi_4(\vec x)' \end{array}\right)
\nonumber \\
&=&\left(\begin{array}{c} - \Psi_3(- \vec x) \\ - \Psi_4(- \vec x) \end{array}
\right) = - \lambda(- \vec x)
\left(\begin{array}{c} \Psi_1(- \vec x) \\ \Psi_2(- \vec x) \end{array}
\right).
\end{eqnarray}
This implies the restriction $\lambda(\vec x) = - \lambda(- \vec x)$.
Rotation invariance and parity limit the anti-Hermitean matrix that
characterizes the boundary condition to
\begin{equation}
\lambda(\vec x) = i \nu \vec \sigma \cdot \vec e_r,
\end{equation}
such that $\vec \sigma \cdot \vec e_r \lambda(\vec x) = i \nu \1$ is indeed
anti-Hermitean. Let us further investigate whether $K$ remains a symmetry in
the finite volume. For this purpose we act with $K$ on the Dirac spinor
$\Psi(\vec x)$, and we ask once more whether the new spinor
$\Psi(\vec x)' = K \Psi(\vec x)$ still obeys the boundary condition. This is
the case only if
\begin{eqnarray}
\lambda(\vec x)
\left(\begin{array}{c} \Psi_1(\vec x)' \\ \Psi_2(\vec x)' \end{array}\right)&=&
\lambda(\vec x) (\vec \sigma \cdot \vec L + \1)
\left(\begin{array}{c} \Psi_1(\vec x) \\ \Psi_2(\vec x) \end{array}\right) =
\left(\begin{array}{c} \Psi_3(\vec x)' \\ \Psi_4(\vec x)' \end{array}\right)
\nonumber \\
&=&- (\vec \sigma \cdot \vec L + \1) \lambda(\vec x)
\left(\begin{array}{c} \Psi_1(\vec x) \\ \Psi_2(\vec x) \end{array}\right),
\end{eqnarray}
which is indeed satisfied since
\begin{equation}
\{\lambda(\vec x),\vec \sigma \cdot \vec L + \1\} =
\{i \nu \vec \sigma \cdot \vec e_r,\vec \sigma \cdot \vec L + \1\} = 0.
\end{equation}
Next, we want to investigate whether the Johnson-Lippmann operator also leaves
the boundary condition invariant. Hence, we now ask whether
$\Psi(\vec x)' = A \Psi(\vec x)$ also obeys the boundary condition, which is
the case only if
\begin{eqnarray}
\lambda(\vec x)
\left(\begin{array}{c} \Psi_1(\vec x)' \\ \Psi_2(\vec x)' \end{array}\right)&=&
\left[\lambda(\vec x) \frac{i}{M c} (\vec \sigma \cdot \vec L + \1)
\vec \sigma \cdot \vec p - \lambda(\vec x) \alpha \vec \sigma \cdot \vec e_r
\right. \nonumber \\
&-&\left. \lambda(\vec x) \frac{i \alpha}{M c R}(\vec \sigma \cdot \vec L + \1)
\lambda(\vec x) \right]
\left(\begin{array}{c} \Psi_1(\vec x) \\ \Psi_2(\vec x) \end{array}\right)
\nonumber \\
&=&\left(\begin{array}{c} \Psi_3(\vec x)' \\ \Psi_4(\vec x)' \end{array}\right)
= \left[\frac{i \alpha}{M c R} (\vec \sigma \cdot \vec L + \1)
\right. \nonumber \\
&-&\left. \frac{i}{M c} (\vec \sigma \cdot \vec L + \1)
\vec \sigma \cdot \vec p \lambda(\vec x) -
\alpha \vec \sigma \cdot \vec e_r \lambda(\vec x) \right]
\left(\begin{array}{c} \Psi_1(\vec x) \\ \Psi_2(\vec x) \end{array}\right).
\nonumber \\ \,
\end{eqnarray}
Using $[\lambda(\vec x),\vec \sigma \cdot \vec e_r] = 0$, this condition would
indeed be satisfied if the following identity holds
\begin{eqnarray}
&&\{\lambda(\vec x),(\vec \sigma \cdot \vec L + \1) \vec \sigma \cdot \vec p\} =
\frac{\alpha}{R} \left(\lambda(\vec x)(\vec \sigma \cdot \vec L + \1)
\lambda(\vec x)
+ \vec \sigma \cdot \vec L + \1 \right) \ \Rightarrow \nonumber \\
&&i \nu \{\vec \sigma \cdot \vec e_r,(\vec \sigma \cdot \vec L + \1)
\vec \sigma \cdot \vec p\} =
\frac{\alpha}{R}(\nu^2 + 1)(\vec \sigma \cdot \vec L + \1).
\end{eqnarray}
It is straightforward to show that
\begin{equation}
i \{\vec \sigma \cdot \vec e_r,(\vec \sigma \cdot \vec L + \1)
\vec \sigma \cdot \vec p\} = (\vec \sigma \cdot \vec L + \1)
i [\vec \sigma \cdot \vec p,\vec \sigma \cdot \vec e_r] =
\frac{2}{R} (\vec \sigma \cdot \vec L + \1)^2,
\end{equation}
which implies that $A$ maintains the boundary condition if
\begin{equation}
\label{Acondition}
\alpha \frac{\nu^2 + 1}{2 \nu} = \vec \sigma \cdot \vec L + \1 = k =
\pm \left(j + \frac{1}{2}\right).
\end{equation}
Obviously, this relation cannot hold as an operator identity. Still, it can
be satisfied for states with an appropriate $k$-value and for specific values
of $\nu$. One might then conclude that, like in the non-relativistic case, a
remnant accidental degeneracy arises for particular states and for specific
values of the self-adjoint extension parameter. However, the situation is more
subtle. In particular, since $\{K,A\} = 0$, an application of $A$ changes the
sign of $k$. This means that, for a fixed value of $\nu$, eq.(\ref{Acondition})
is no longer satisfied for $-k$, and thus another application of $A$ will
inevitably lead out of the domain of $H$. Since repeated applications of $H$ or
$K$, on the other hand, leave the wave function inside the domain,
eq.(\ref{susyH}), which relates $A^2$ to $H^2$ and $K^2$ in the infinite volume,
is no longer satisfied in a finite volume. In fact, $H$ and $A$ no longer
commute as operators restricted to the domain of $H$. This implies that $A$ no
longer generates an accidental symmetry, and the corresponding degeneracy is
lifted in a finite volume.

\begin{figure}[tbh]
\begin{center}
\epsfig{file=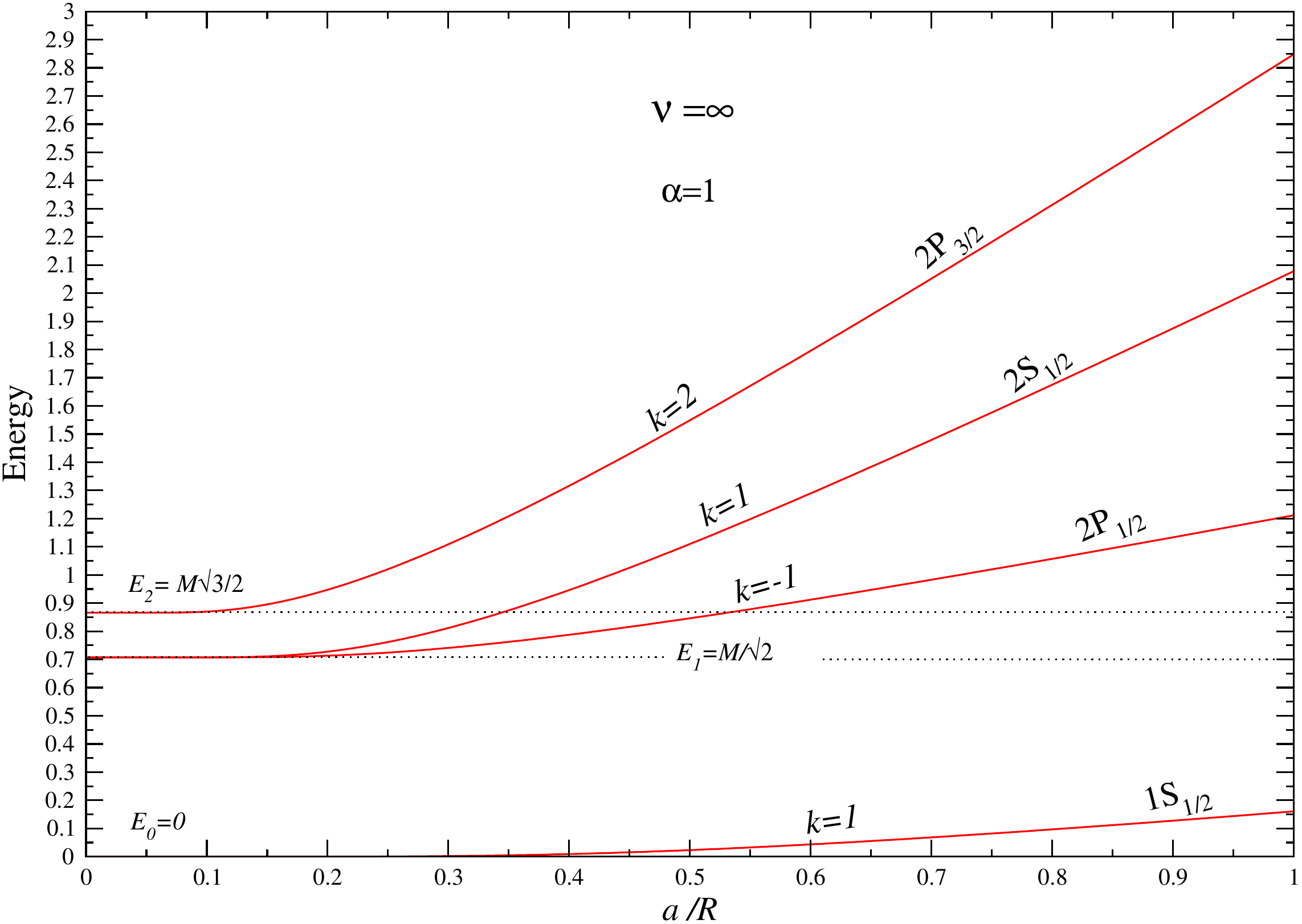,width=10cm} \\
\epsfig{file=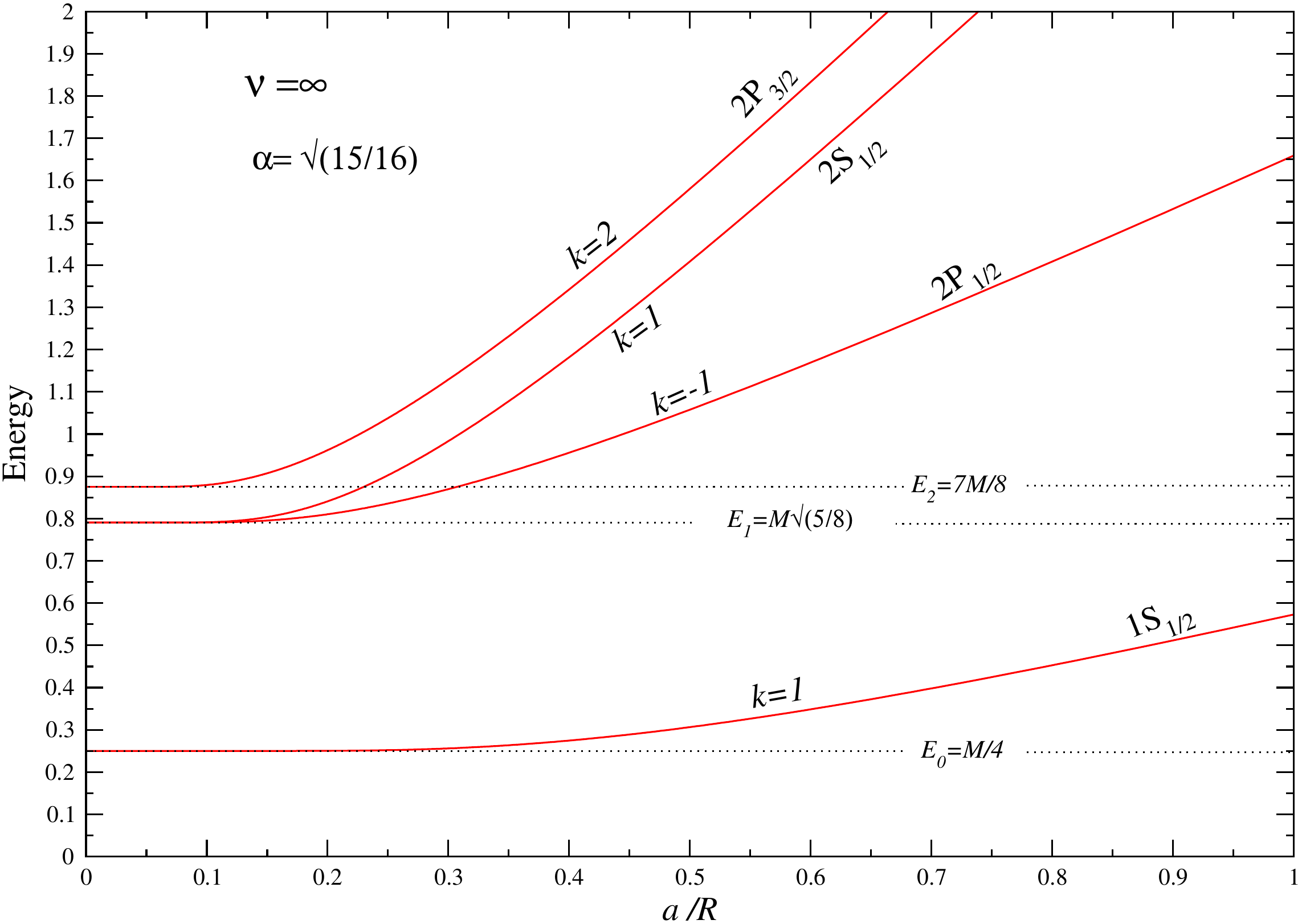,width=10cm}
\end{center}
\caption{\it Spectrum of the Dirac hydrogen atom centered in a spherical
cavity with the standard Dirichlet boundary condition (i.e.\ $\nu = \infty$)
for $\alpha = 1$ (top) and for $\alpha = \sqrt{15/16}$ (bottom), as a function
of $a/R$. The energies of states with quantum numbers $n = 1$ and 2 are given
in units of $M$. The dotted lines represent the spectrum of the infinite
system.}
\label{specrelinf}
\end{figure}

\begin{figure}[tbh]
\begin{center}
\epsfig{file=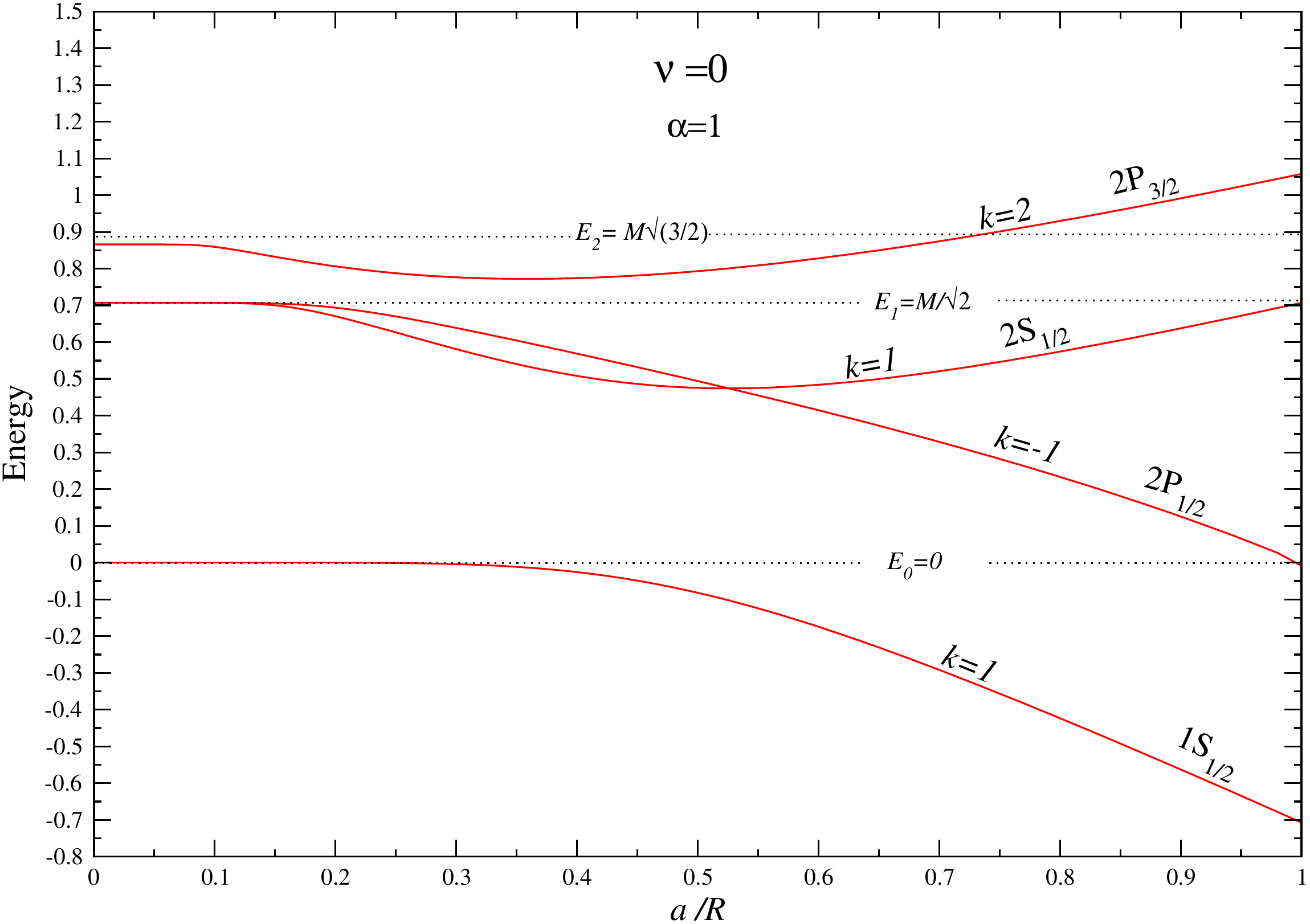,width=10cm} \\
\epsfig{file=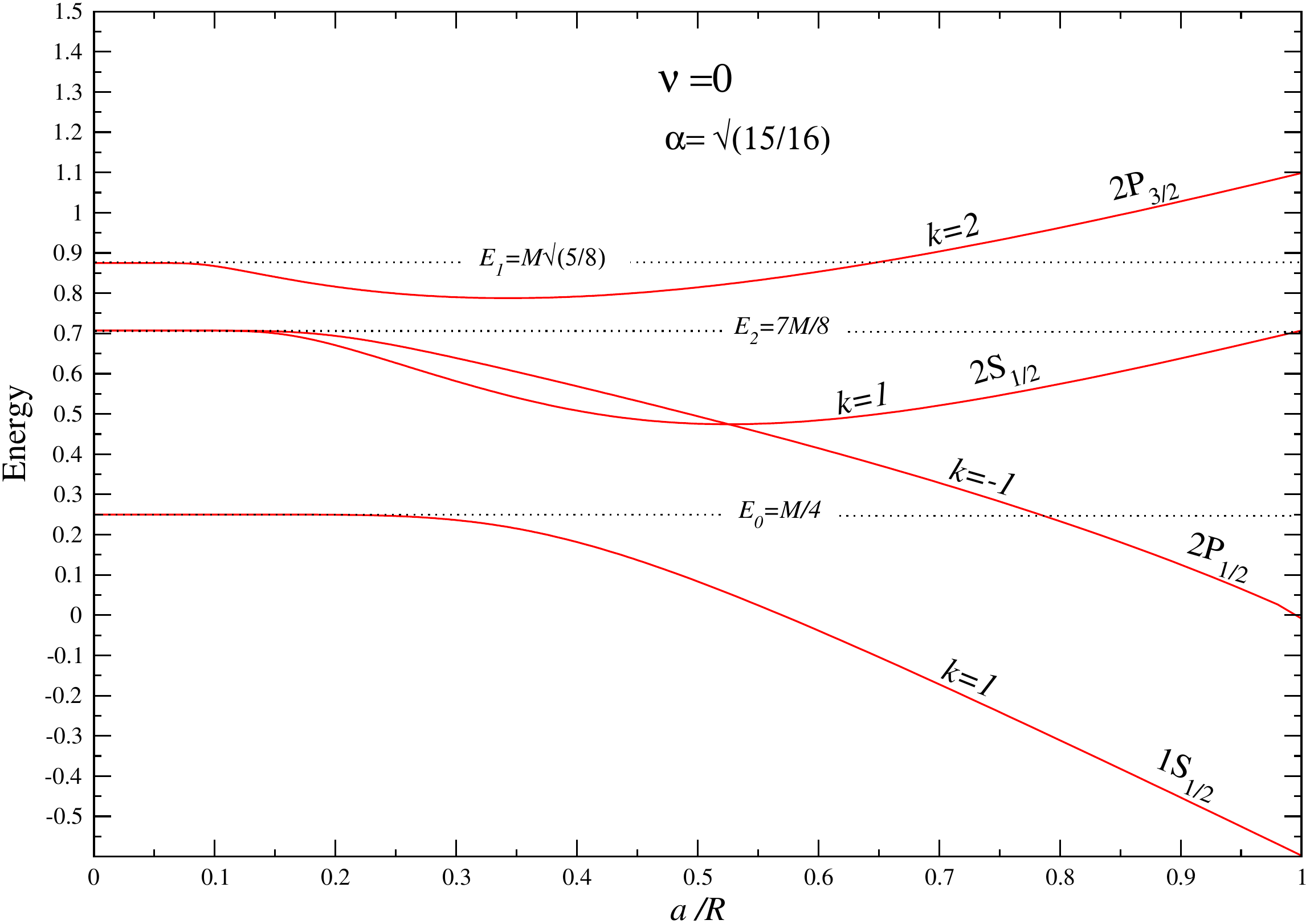,width=10cm}
\end{center}
\caption{\it Spectrum of the Dirac hydrogen atom centered in a spherical
cavity with Neumann boundary condition (i.e.\ $\nu = 0$) for $\alpha = 1$
(top) and for $\alpha = \sqrt{15/16}$ (bottom), as a function of $a/R$. The
energies of states with quantum numbers $n = 1$ and 2 are given in units of
$M$. The dotted lines represent the spectrum of the infinite system.}
\label{specrel0}
\end{figure}

Let us now investigate the spectrum of relativistic hydrogen in a spherical
cavity in more detail. First we make a separation ansatz for the wave function
\begin{equation}
\Psi(\vec x) = \left(\begin{array}{c} \Psi_1(\vec x) \\ \Psi_2(\vec x) \\
\Psi_3(\vec x) \\ \Psi_4(\vec x) \end{array} \right) =
\left(\begin{array}{c} \psi_A(r) {\cal Y}_{jj_3l_A}(\theta,\varphi) \\
i \psi_B(r) {\cal Y}_{jj_3l_B}(\theta,\varphi) \end{array} \right).
\end{equation}
For $k = j + \frac{1}{2}$ we have $l_A = j - \frac{1}{2}$ and
$l_B = j + \frac{1}{2}$, while for $k = - (j + \frac{1}{2})$,
$l_A = j + \frac{1}{2}$ and $l_B = j - \frac{1}{2}$. For $j = l + \frac{1}{2}$
the spin-angular functions are given by
\begin{equation}
{\cal Y}_{jj_3l}(\theta,\varphi) = \sqrt{\frac{l + j_3 + \frac{1}{2}}{2 l + 1}}
Y_{l,j_3 - \frac{1}{2}}(\theta,\varphi)
\left(\begin{array}{c} 1 \\ 0 \end{array} \right) +
\sqrt{\frac{l - j_3 + \frac{1}{2}}{2 l + 1}}
Y_{l,j_3 + \frac{1}{2}}(\theta,\varphi)
\left(\begin{array}{c} 0 \\ 1 \end{array} \right),
\end{equation}
while for $j = l - \frac{1}{2}$
\begin{equation}
{\cal Y}_{jj_3l}(\theta,\varphi) =
- \sqrt{\frac{l - j_3 + \frac{1}{2}}{2 l + 1}}
Y_{l,j_3 - \frac{1}{2}}(\theta,\varphi)
\left(\begin{array}{c} 1 \\ 0 \end{array} \right) +
\sqrt{\frac{l + j_3 + \frac{1}{2}}{2 l + 1}}
Y_{l,j_3 + \frac{1}{2}}(\theta,\varphi)
\left(\begin{array}{c} 0 \\ 1 \end{array} \right).
\end{equation}
Here $Y_{lm}(\theta,\varphi)$ are the usual spherical harmonics. The radial
equations then take the form
\begin{eqnarray}
&&\left(M c^2 - \frac{e^2}{r}\right) \psi_A(r) -
c \left(\p_r + \frac{1 + k}{r}\right) \psi_B(r) = E \psi_A(r), \nonumber \\
&&\left(- M c^2 - \frac{e^2}{r}\right) \psi_B(r) +
c \left(\p_r + \frac{1 - k}{r}\right) \psi_A(r) = E \psi_B(r).
\end{eqnarray}
The cavity boundary condition is given by
\begin{eqnarray}
&&i \nu \vec e_r \cdot \vec \sigma \psi_A(R) {\cal Y}_{jj_3l_A}(\theta,\varphi) =
- i \nu \psi_A(R) {\cal Y}_{jj_3l_B}(\theta,\varphi) =
i \psi_B(R) {\cal Y}_{jj_3l_B}(\theta,\varphi) \ \Rightarrow \nonumber \\
&&\nu \psi_A(R) = - \psi_B(R).
\end{eqnarray}
The resulting finite volume spectra for $\nu = \infty$ (i.e.\ Dirichlet
boundary conditions) and for $\nu = 0$ (i.e.\ Neumann boundary conditions)
are illustrated in Figs.\ \ref{specrelinf} and \ref{specrel0}, respectively.
In order to make the effects easily visible we have chosen very large
unphysical values of $\alpha$.
In both cases, we see that the states 2S$_{1/2}$ and 2P$_{1/2}$, which are
accidentally degenerate in the infinite volume, are split in the finite cavity.
With Dirichlet boundary conditions ($\nu = \infty$) the energies increase with
decreasing cavity radius, while for Neumann boundary conditions ($\nu = 0$) the
energy of some states decreases.

\begin{figure}[tbh]
\begin{center}
\epsfig{file=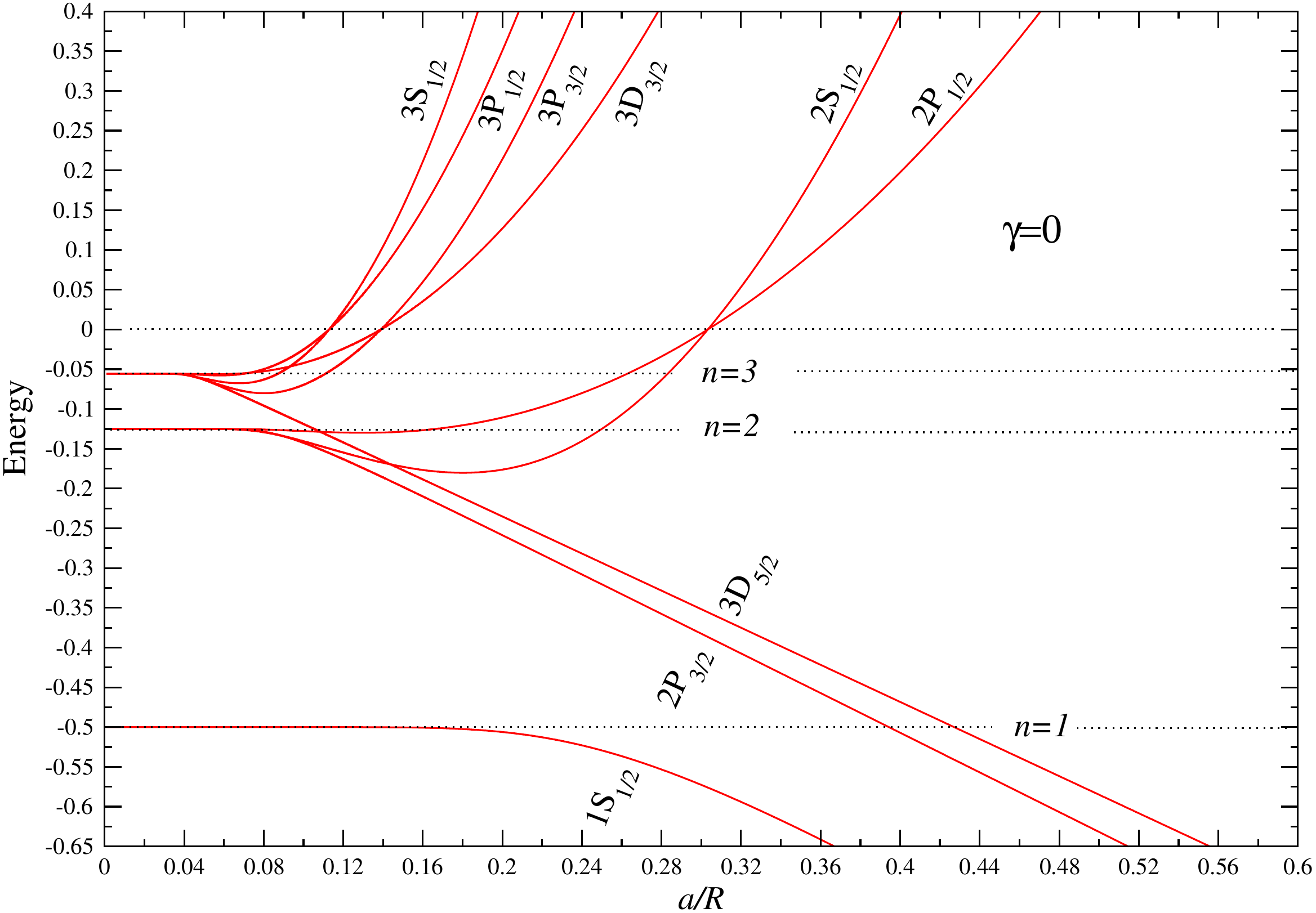,width=10cm}
\end{center}
\caption{\it Spectrum of the Pauli hydrogen atom centered in a spherical
cavity with Neumann boundary condition (i.e.\ $\gamma = 0$) as a function of
$a/R$. The energies of states with quantum numbers $n = 1, 2, 3$ are given in
units of $M e^4$. The dotted lines represent the spectrum of the infinite
system.}
\label{Paulispec0}
\end{figure}

\subsection{The Non-relativistic Pauli Atom in a Spherical Cavity}

Finally, we proceed to the non-relativistic limit of the Dirac equation, which
leads us to the Pauli equation
\begin{equation}
H = M c^2 + \frac{\left(\vec p c + e \vec A(\vec x)\right)^2}{2 M c^2} -
e \Phi(\vec x) + \mu \vec \sigma \cdot \vec B(\vec x).
\end{equation}
Since we have no external magnetic field, at least to the leading order we are
working at, spin decouples and the Pauli equation reduces to two copies of the
Schr\"odinger equation. In particular, we have neglected the sub-leading
spin-orbit couplings in the Hamiltonian. Still, as we discussed in Subsection
3.3, the spin explicitly enters the boundary condition
\begin{equation}
\label{Paulibc}
\gamma(\vec x) \Psi(\vec x) + \vec n(\vec x) \cdot
\left[\vec \nabla  \Psi(\vec x) -
i \vec \sigma \times \vec \nabla \Psi(\vec x)\right] = 0,
\end{equation}
which thus leads to a spin-orbit coupling induced by the cavity wall. The
Hermitean matrix $\gamma(\vec x)$, which results from the anti-Hermitean matrix
$\lambda(\vec x)$ in the non-relativistic limit, is given by
\begin{equation}
\gamma(\vec x) = - 2 M c i \vec n(\vec x) \cdot \vec \sigma \lambda(\vec x) =
2 M c \nu \1.
\end{equation}
For $j = l + \frac{1}{2}$ the boundary condition of eq.(\ref{Paulibc}) reduces
to
\begin{equation}
\label{bcj+}
\gamma \psi_{nl}(R) + \p_r \psi_{nl}(R) - \frac{l}{R} \psi_{nl}(R) = 0,
\end{equation}
while for $j = l - \frac{1}{2}$
\begin{equation}
\label{bcj-}
\gamma \psi_{nl}(R) + \p_r \psi_{nl}(R) + \frac{l+1}{R} \psi_{nl}(R) = 0,
\end{equation}
The spectrum of the Pauli equation for Neumann boundary conditions
(i.e.\ $\gamma = 0$) is illustrated in Fig.\ \ref{Paulispec0}. It differs from
the spectrum of Fig.\ \ref{spec0} because now the angular momentum enters the
boundary condition.

Let us now address the question of remnant accidental degeneracies that persist
even in a finite volume. As we discussed in Subsection 4.1, for the
Schr\"odinger atom these exist for $R = (l + 1)(l + 2)a$ and $\gamma = 2/R$ or
$\infty$, because, in these cases, two applications of the Runge-Lenz vector
lead back into the domain of the Hamiltonian. For the Dirac atom in a finite
volume, on the other hand, repeated applications of the Johnson-Lippmann
operator do not lead back into the domain of the Hamiltonian, and thus the
accidental degeneracy is lifted in a finite volume (c.f.\ Subsection 4.2).

\begin{figure}[tbh]
\begin{center}
\epsfig{file=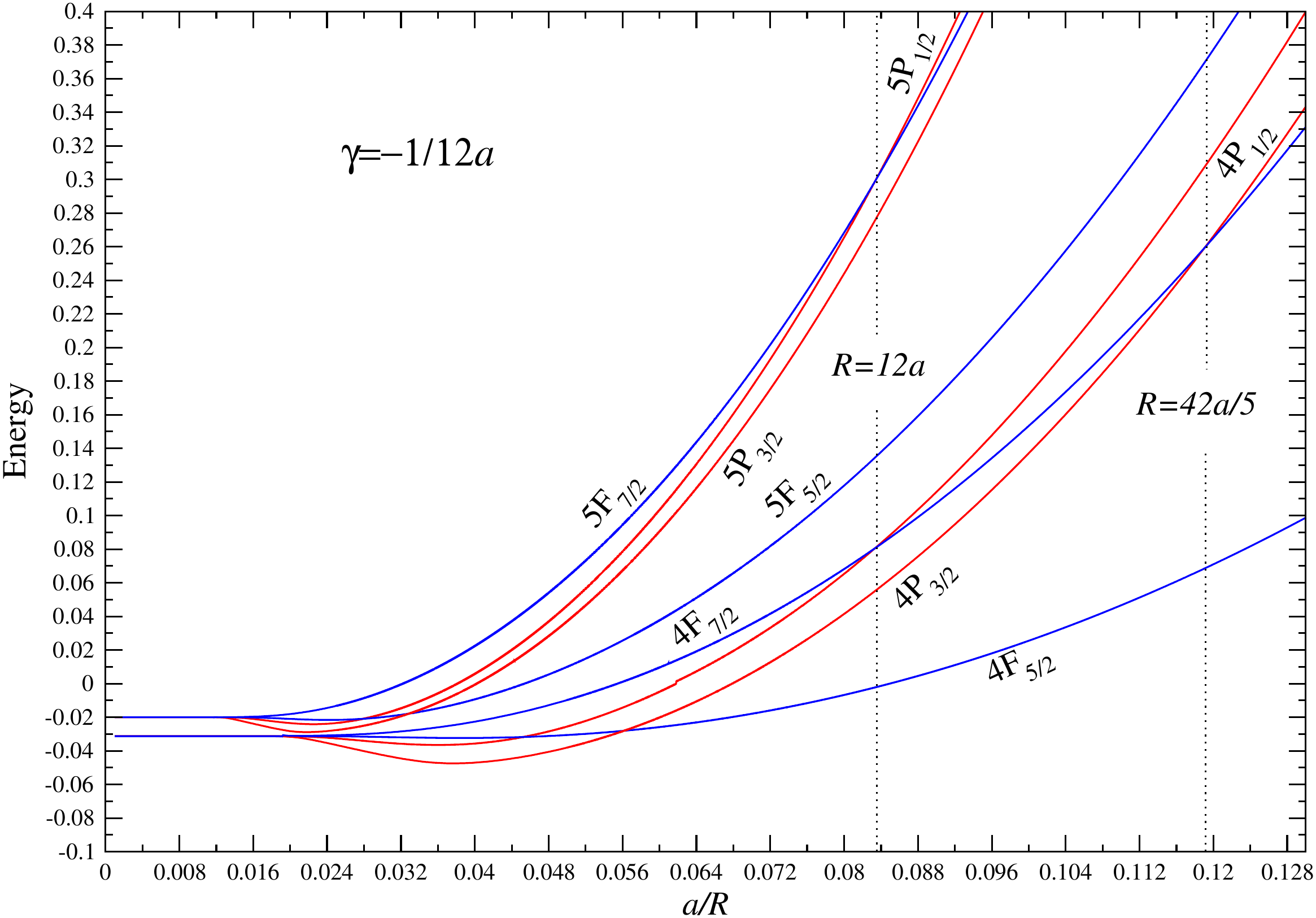,width=7.25cm}
\epsfig{file=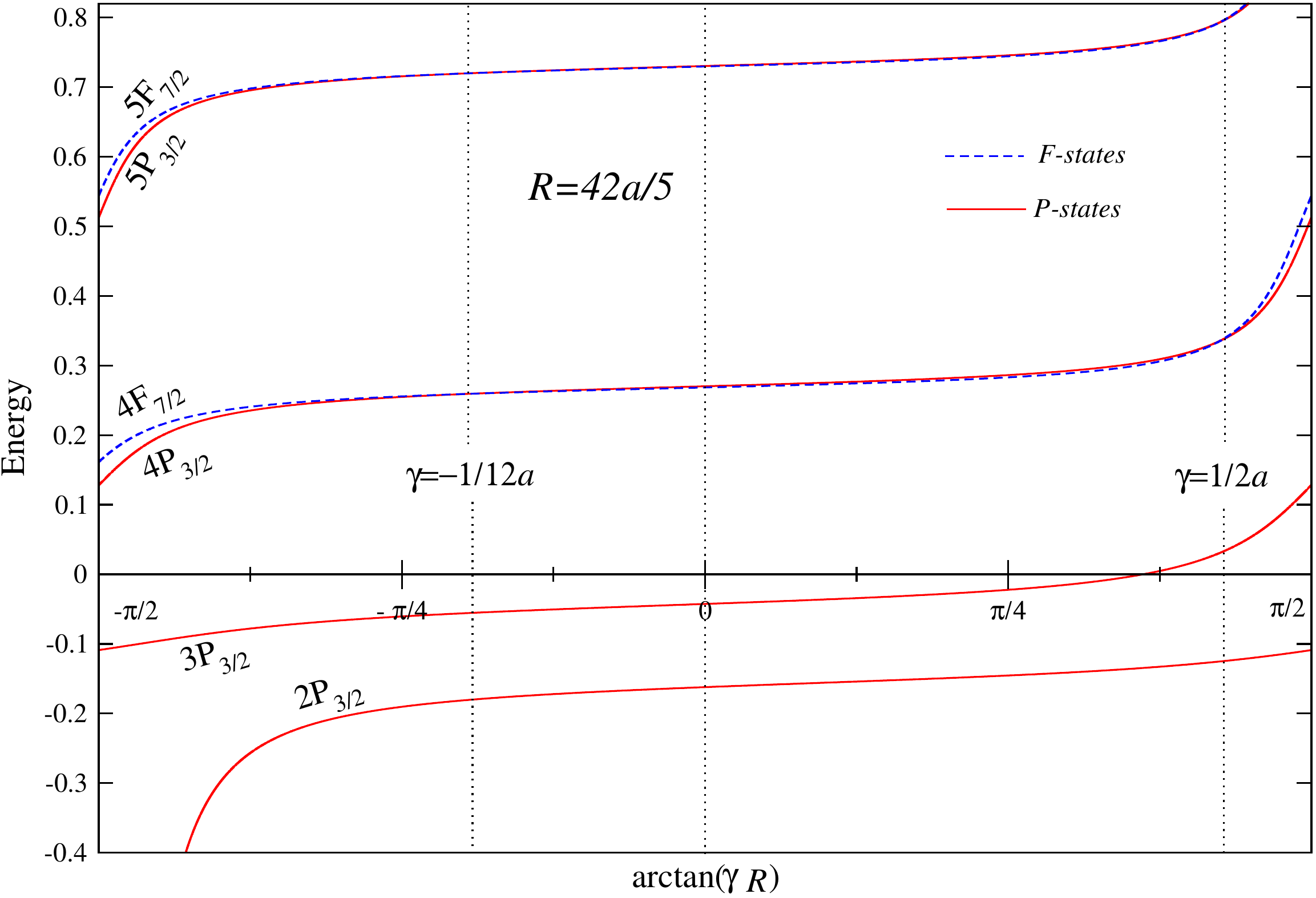,width=7.25cm} \vskip0.5cm
\epsfig{file=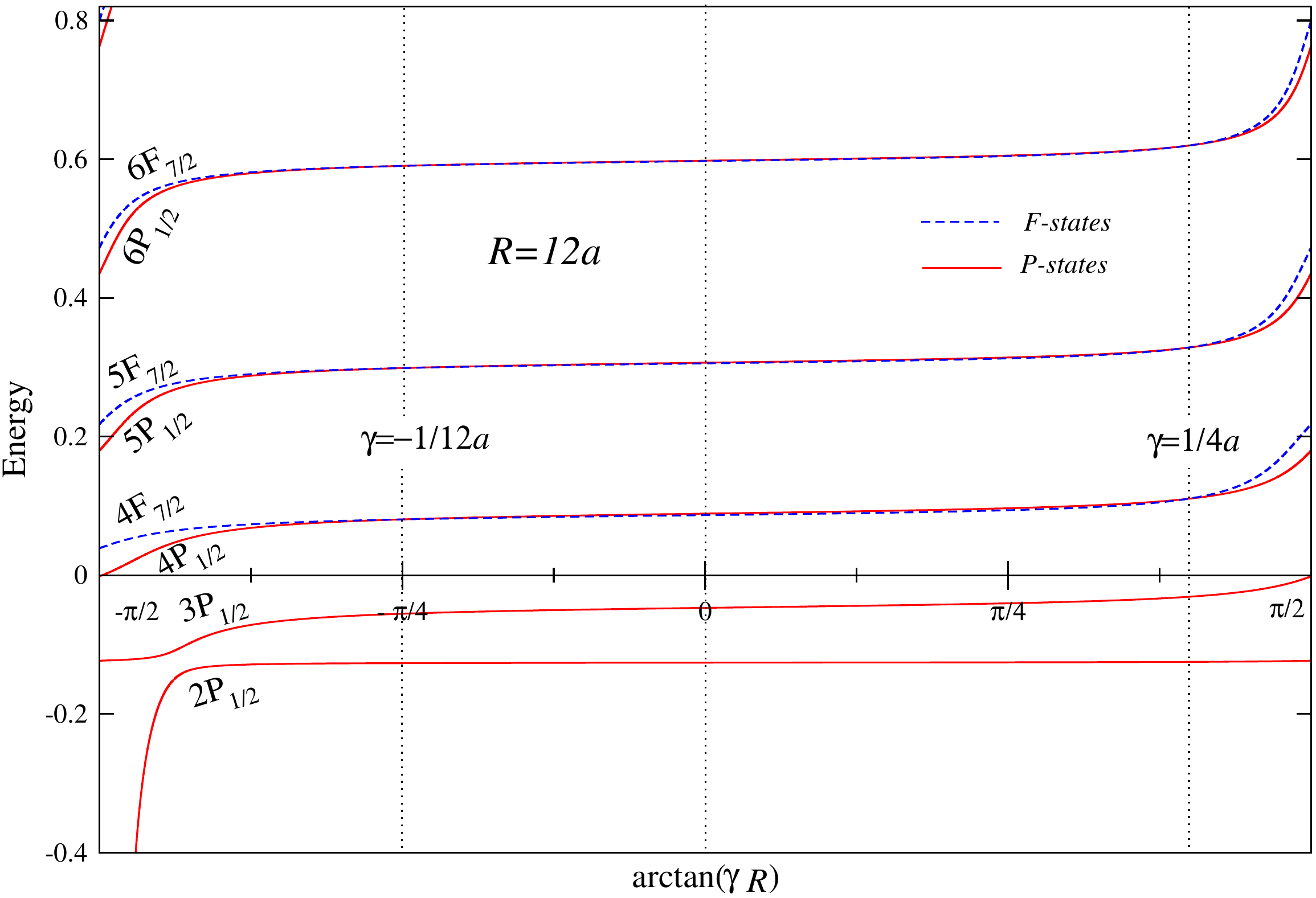,width=7.25cm}
\epsfig{file=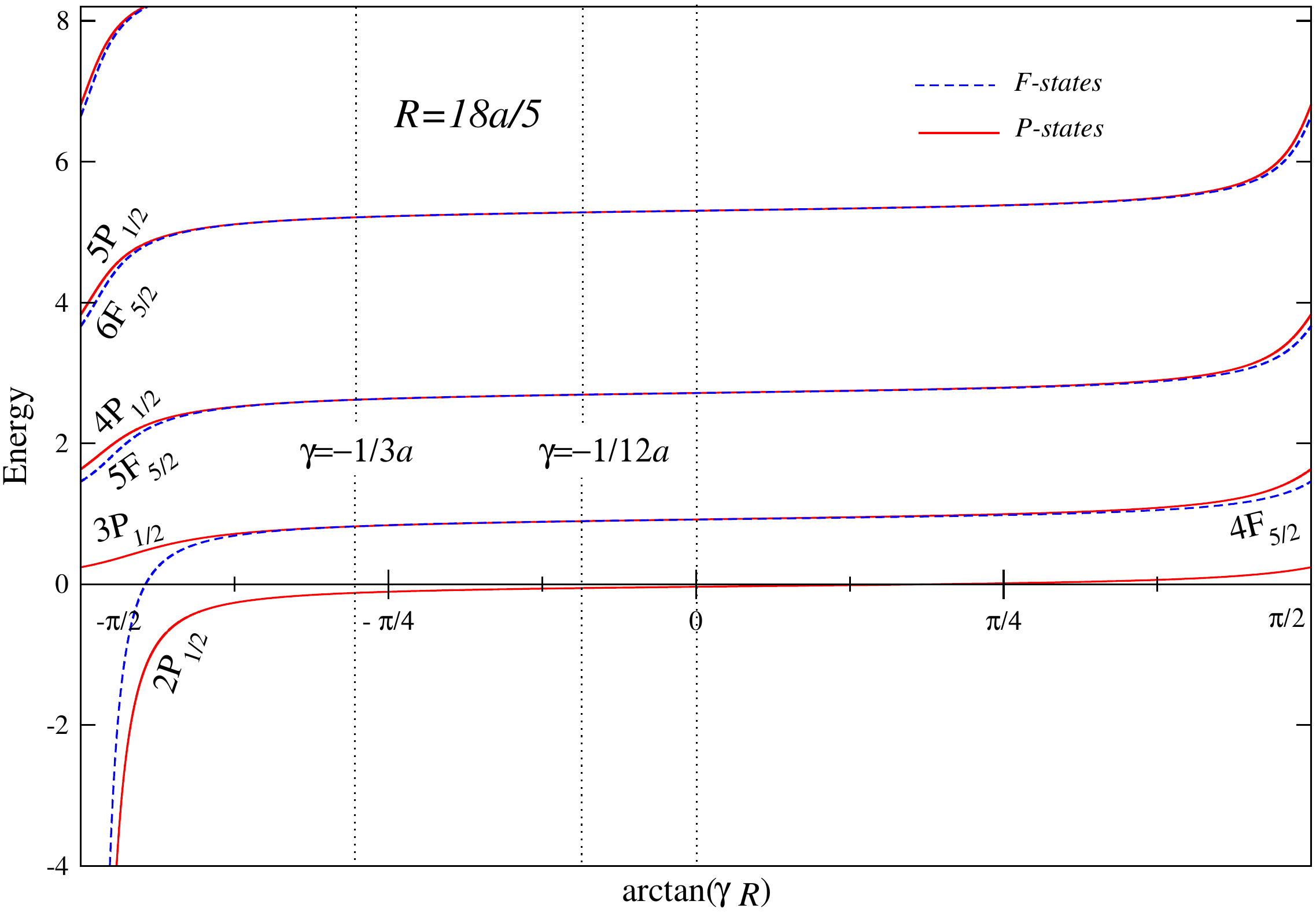,width=7.25cm}
\end{center}
\caption{\it Top left: Energy of P- and F-states with $n = 4$ and 5 for the
Pauli hydrogen atom in a spherical cavity with $\gamma = - \frac{1}{12a}$ as a
function of $a/R$. There are accidental degeneracies (indicated by the vertical
dashed lines) of the 4P$_{3/2}$ with the 4F$_{7/2}$ for $R = \frac{42}{5} a$, as
well as of the 4P$_{1/2}$ with the 4F$_{7/2}$ and of the 5P$_{1/2}$ with the
5F$_{7/2}$ states for $R = 12 a$. Top right and bottom: Energy of P- and F-states
for the Pauli hydrogen atom in a spherical cavity as a function of
$\arctan(\gamma R)$ at fixed cavity radius: $R = \frac{42}{5} a$ (top right),
$R = 12 a$ (bottom left), and $R = \frac{18}{5} a$ (bottom right): The energies
of the high-lying P- and F-states are very similar for all values of $\gamma$,
but are identical only for the special values indicated by the vertical dashed
lines. The energy is given in units of $M e^4$.}
\label{relaccidental}
\end{figure}

As we will see now, just as for the Schr\"odinger atom, for the Pauli atom some
accidental degeneracies persist in a finite cavity. We start again from
eq.(\ref{chieq}), take its derivative with respect to $r$, and then
eliminate $\partial_r^2 \psi_{nl}(r)$ by using the radial Schr\"odinger
equation (\ref{radialSE}). We then impose the boundary condition eq.(\ref{bcj+})
when $j = l + \frac{1}{2}$, or eq.(\ref{bcj-}) when $j = l - \frac{1}{2}$.
Finally, we ask under what circumstances the wave function $\chi_{n,l+2}(r)$ of
eq.(\ref{chieq}) satisfies the corresponding boundary condition
\begin{equation}
\gamma \chi_{n,l+2}(R) + \p_r \chi_{n,l+2}(R) - \frac{l+2}{R} \chi_{n,l+2}(R) = 0,
\end{equation}
when $j' = l + 2 + \frac{1}{2} = l + \frac{5}{2}$, or
\begin{equation}
\gamma \chi_{n,l+2}(R) + \p_r \chi_{n,l+2}(R) + \frac{l+3}{R} \chi_{n,l+2}(R) = 0,
\end{equation}
when $j' = l + 2 - \frac{1}{2} = l + \frac{3}{2}$, for all values of the
energy $E$. For $j = l + \frac{1}{2}$ and $j' = l + \frac{5}{2}$ this turns out
to be the case for
\begin{equation}
\label{magic1}
R = \frac{(l + 1)(l + 2)(2l + 5) a}{2 l + 3}, \quad
\gamma = - \frac{1}{2(l + 1)(l + 2) a}, \ \mbox{or} \
\gamma = \frac{1}{(l + 1)a}.
\end{equation}
For $j = l + \frac{1}{2}$ and $j' = l + \frac{3}{2}$, on the other hand, the
conditions can be satisfied only when $e^2 = 0$ and $\gamma = 0$. For
$j = l - \frac{1}{2}$ and $j' = l + \frac{5}{2}$ one obtains
\begin{equation}
\label{magic2}
R = 2(l + 1)(l + 2)a, \quad
\gamma = - \frac{1}{2(l + 1)(l + 2) a} = - \frac{1}{R}, \ \mbox{or} \
\gamma = \frac{3}{2(l + 1)(l + 2) a} = \frac{3}{R},
\end{equation}
and, finally, for $j = l - \frac{1}{2}$ and $j' = l + \frac{3}{2}$ one finds
\footnote{We like to thank D.\ Banerjee for valuable help with deriving
eqs.(\ref{magic1}), (\ref{magic2}), and (\ref{magic3}).}
\begin{equation}
\label{magic3}
R = \frac{(l + 1)(l + 2)(2l + 1) a}{2 l + 3}, \quad
\gamma = - \frac{1}{2(l + 1)(l + 2) a},  \ \mbox{or} \
\gamma = - \frac{1}{(l + 2)a}.
\end{equation}
The remnant accidental symmetries for the Pauli hydrogen atom are illustrated
in Fig.\ \ref{relaccidental}.

\section{Conclusions}

We have investigated the accidental symmetry in the relativistic hydrogen atom
confined to a spherical cavity with perfectly reflecting boundary conditions.
In the infinite volume the Johnson-Lippman operator $A$, which is the
relativistic analog of the Runge-Lenz vector $\vec R$, commutes with the
Hamiltonian, thus leading to accidental degeneracies in the energy spectrum.
When the system is placed in a finite volume, the accidental symmetry is lifted.
This is because a repeated application of the Johnson-Lippman operator leads out
of the domain of the Hamiltonian. In the non-relativistic case, for specific
values of the cavity radius $R$ and the self-adjoint extension parameter
$\gamma$, a repeated application of $\vec R$ leads back into the domain of the
Hamiltonian, and thus some accidental degeneracy persists even in a finite
volume. Because repeated applications of $A$ do not lead back into the domain
of the Hamiltonian, this is not the case for the Dirac equation. Interestingly,
in the Pauli equation with spin, which results as the non-relativistic limit of
the Dirac equation, the boundary conditions induce non-trivial spin-orbit
couplings. Remarkably, in this case, for specific values of the cavity radius
$R$ and the self-adjoint extension parameter $\gamma$, a repeated application
of $\vec R$ again leads back into the domain of the Hamiltonian, and thus some
accidental degeneracy persists in a finite volume.

Our investigation shows that the subtle issues of Hermiticity versus
self-ad\-joint\-ness and the domain structure of quantum mechanical Hamiltonians
hold the key to understanding in detail what happens to a hydrogen atom when
it is confined in a spherical cavity. The same is true for other accidental
symmetries, for example, for the Landau level problem \cite{AlH09}, for a
particle on a cone \cite{AlH12a}, or for the harmonic oscillator
\cite{AlH12a,AlH13}. This underscores that the theory of self-adjoint extensions
is not just a mathematical curiosity, but of great importance for understanding
the physics of confined systems.

\section*{Acknowledgments}

This publication was made possible by the NPRP grant \# NPRP 5 - 261-1-054 from
the Qatar National Research Fund (a member of the Qatar Foundation). The
statements made herein are solely the responsibility of the authors.

\end{document}